\documentclass{natureprintstyle}
\usepackage{graphicx,amsmath}
\usepackage{lscape}
\usepackage{color}
\usepackage{url}
\usepackage{ulem}
\usepackage{multirow}
\usepackage{caption}
\usepackage[square,sort,comma,numbers]{natbib}
\usepackage{multibib}
\usepackage{graphics,graphicx,amsmath,amssymb,xcolor}
\usepackage[colorlinks=true,linkcolor=blue,citecolor=blue,urlcolor=blue]{hyperref}
\usepackage{amssymb}
\newcites{Extra}{Additional References}
\bibliographystyleExtra{naturemag}

\usepackage{astjnlabbrev-nature}

\title{Discovery of extreme Quasi-Periodic Eruptions in a newly accreting massive black hole}

\author{Lorena Hernández-García$^{1,2,3}$, 
Joheen Chakraborty$^{4}$, 
Paula Sánchez-Sáez$^{5}$, 
Claudio Ricci$^{6,7}$,
Jorge Cuadra$^{8,1}$,
Barry McKernan$^{9,10,11}$,
K.E. Saavik Ford$^{9,10,11}$,
Patricia Arévalo$^{3,1,2}$, 
Arne Rau$^{12}$,
Riccardo Arcodia$^{4}$, 
Erin Kara$^{4}$, 
Zhu Liu$^{12}$,
Andrea Merloni$^{12}$,
Gabriele Bruni$^{13}$,
Adelle Goodwin$^{14}$,
Zaven Arzoumanian$^{15}$,
Roberto J. Assef$^{6}$,
Pietro Baldini$^{12}$,
Amelia Bayo$^{5}$, 
Franz E. Bauer$^{16,17,2,18}$, 
Santiago Bernal$^{3,1}$,
Murray Brightman$^{19}$,
Gabriela Calistro Rivera$^{5,20}$,
Keith Gendreau$^{15}$,
David Homan$^{21}$,
Mirko Krumpe$^{21}$,
Paulina Lira$^{22,1}$,
Mary Loli Martínez-Aldama$^{23,1,2}$,
Mara Salvato$^{12}$, and
Belén Sotomayor$^{3}$
}

\newcommand\arcsec{\mbox{$^{\prime\prime}$}}%
\newcommand{\nicer}{\textit{NICER}}
\newcommand{\xmm}{{\it XMM-Newton}}
\newcommand{\swift}{{\it Swift}}
\usepackage{float}

\DeclareUnicodeCharacter{2212}{-}

\begin{document}

\maketitle

\begin{affiliations}
 \item Millennium Nucleus on Transversal Research and Technology to Explore Supermassive Black Holes (TITANS), Gran Breta\~na 1111, Playa Ancha, Valpara\'iso, Chile
 
 \item Millennium Institute of Astrophysics (MAS), Nuncio Monseñor Sótero Sanz 100, Providencia, Santiago, Chile
 
 \item Instituto de F\'isica y Astronom\'ia, Facultad de Ciencias, Universidad de Valpara\'iso, Gran Breta\~na 1111, Playa Ancha, Valpara\'iso, Chile
 
 \item MIT Kavli Institute for Astrophysics and Space Research, Massachusetts Institute of Technology, Cambridge, MA 02139, USA
 
\item European Southern Observatory, Karl-Schwarzschild-Strasse 2, 85748 Garching bei München, Germany

\item Instituto de Estudios Astrof\'isicos, Facultad de Ingenier\'ia y Ciencias, Universidad Diego Portales, Av. Ej\'ercito Libertador 441, Santiago, Chile

\item Kavli Institute for Astronomy and Astrophysics, Peking University, Beijing 100871, People’s Republic of China

\item Departamento de Ciencias, Facultad de Artes Liberales, Universidad Adolfo Ibáñez, Av.\ Padre Hurtado 750, Viña del Mar, Chile

\item Dept of Science, BMCC, City University of New York, NY 10007, USA

\item  Graduate Center, City University of New York, 365 5th Avenue, NY 10016, USA

\item  Dept. of Astrophysics, American Museum of Natural History,  NY 10024, USA

\item Max-Planck-Institut f\"ur Extraterrestrische Physik, Gie{\ss}enbachstra{\ss}e, 85748 Garching, Germany

\item INAF -- Istituto di Astrofisica e Planetologia Spaziali, via del Fosso del Cavaliere 100, Roma, 00133, Italy

\item International Centre for Radio Astronomy Research – Curtin University, GPO Box U1987, Perth, WA 6845, Australia

\item X-ray Astrophysics Laboratory, Code 662, NASA Goddard Space Flight Center, Greenbelt, MD 20771, USA

\item Instituto de Astrof{\'{\i}}sica, Facultad de F{\'{i}}sica, Pontificia Universidad Cat{\'{o}}lica de Chile, Campus San Joaquín, Av. Vicuña Mackenna 4860, Macul Santiago, Chile, 7820436

\item Centro de Astroingenier{\'{\i}}a, Facultad de F{\'{i}}sica, Pontificia Universidad Cat{\'{o}}lica de Chile, Campus San Joaquín, Av. Vicuña Mackenna 4860, Macul Santiago, Chile, 7820436

\item Space Science Institute, 4750 Walnut Street, Suite 205, Boulder, Colorado 80301, USA

\item Cahill Center for Astrophysics, California Institute of Technology, 1216 East California Boulevard, Pasadena, CA 91125, USA

\item German Aerospace Center (DLR), Institute of Communications and Navigation, Wessling, Germany

\item Leibniz-Institut f\"ur Astrophysik Potsdam (AIP), An der Sternwarte 16, 14482 Potsdam, Germany

\item Departamento de Astronomía, Universidad de Chile, Casilla 36D, Santiago, Chile

\item Astronomy Department, Universidad de Concepci\'on, Barrio Universitario S/N, Concepci\'on 4030000, Chile

\end{affiliations}

\begin{abstract}

Quasi-periodic eruptions (QPEs) are rapid, recurring X-ray bursts from supermassive black holes, believed to result from interactions between accretion disks and surrounding matter. The galaxy SDSS1335+0728, previously stable for two decades, exhibited an increase in optical brightness in December 2019, followed by persistent Active Galactic Nucleus (AGN)-like variability for 5 years, suggesting the activation of a $\sim$10$^6\;M_\odot$ black hole. From February 2024, X-ray emission has been detected, revealing extreme $\sim$4.5-day QPEs with the highest fluxes and amplitudes, longest time scales, largest integrated energies, and a $\sim$25-day super-period. Low-significance UV variations are reported for the first time in a QPE host, likely related to the long timescales and large radii from which the emission originates. This discovery broadens the possible formation channels for QPEs, suggesting they are not linked solely to tidal disruption events but more generally to newly formed accretion flows, which we are witnessing in real time in a turn-on AGN candidate.

\end{abstract}

\begin{figure*}
 \centering
 \includegraphics[width=\textwidth]{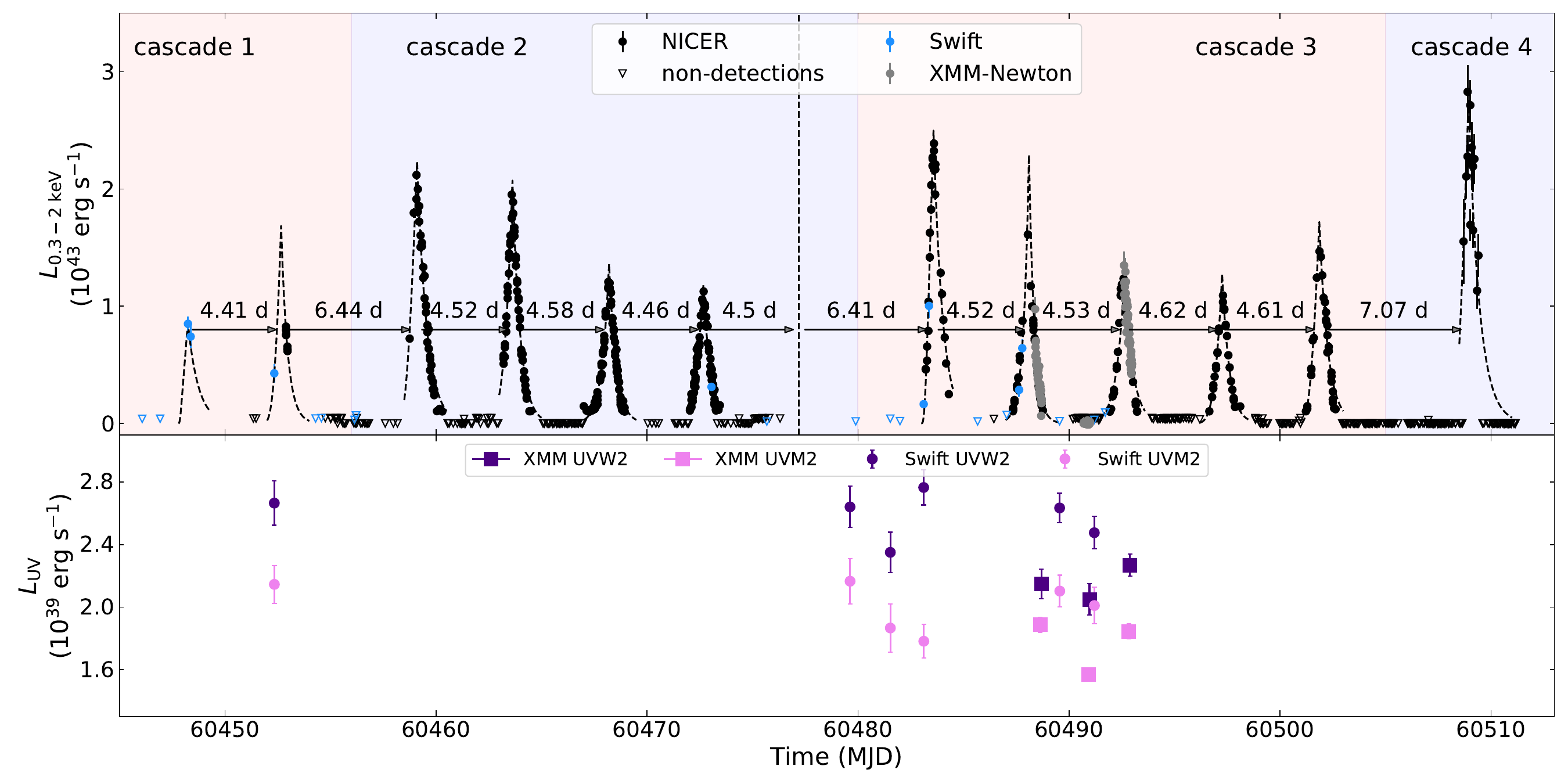}
 \caption{(Top panel): X-ray light curve of Ansky (using \textit{NICER} XTI/\textit{Swift} XRT/\textit{XMM-Newton} EPIC pn) from May 15-July 22, 2024, showing QPEs with a typical peak-to-peak recurrence time of $\sim$4.5~d. Shaded regions correspond to super-periods of 5 consecutive QPEs, which we term ``cascades'' due to the evolving peak luminosity. Cascades are separated by a longer separation than most QPEs. The dashed lines are exponential rise and decay profiles fit to each QPE, as described in Methods. The vertical dashed line represents an expected peak in a time range when no observations were carried out. (Bottom panel): UV light curves from the \textit{XMM-Newton}/OM (squares) and \textit{Swift}/UVOT (circles) instruments in the UVW2 (purple) and UVM2 (pink) bands.}
\label{fig:lc}
\end{figure*}

The galaxy SDSS\,J133519.91+072807.4 (z=0.024), which had exhibited no prior optical variations during the preceding two decades, began showing significant nuclear variability in the Zwicky Transient Facility (ZTF) alert stream from December 2019 (originally as ZTF19acnskyy and later as ZTF22abyhaut). 
In this manuscript, we refer to this event as ``Ansky'', which comes from the first object ID reported by ZTF.
Multi-wavelength photometric and spectroscopic follow-up analysis \cite{Sanchez24} found that: 
(a) the UV flux increased by a factor of four between 2004 (GALEX; \cite{Martin05}) and 2021 (\textit{Swift}/UVOT; \cite{Burrows05})
(b) since June 2022, the mid-infrared flux has more than doubled, and the W1-W2 WISE \cite{Wright10,Mainzer14} color has become redder; 
(c) the [O\,{\sc iii}] line flux increased by a factor of $>\sim$ 5 in the last two years, implying a compact ($\sim$1 pc) narrow line emitting region, with narrow emission line ratio evolution now consistent with a more energetic ionizing continuum typical of AGN;
(d) no broad emission lines have been detected as yet; 
(e) the optical light curve shows a decay profile much flatter than either the theoretical or observed ranges for known tidal disruption events (TDEs; \cite{vanVelzen20}), and
(f) since February 2024, the source has begun showing X-ray emission.  
This behaviour, and the known properties of the host, suggest that Ansky hosts a $\sim$10$^6\;M_\odot$ black hole, as estimated from the galaxy stellar mass \cite{Reines15}, and is currently in the process of `turning on'. 

\noindent We focus here on the nature of the recently detected X-ray emission. We complement this study with UV and optical observations. Radio observations are also reported (Methods), however no emission is detected in the 
150 MHz, 0.8 GHz, 1.4 GHz, 3 GHz, 5.5 GHz and 9 GHz bands.

Archival X-ray observations show that the source was not detected by the ROSAT all-sky survey (upper limit of $F=1.6\times 10^{-13}$ erg cm$^{−2}$ s$^{−1}$ or $L=2.1\times 10^{41}$ erg s$^{−1}$ in the 0.2-2 keV energy band) nor by the German eROSITA All-Sky Survey (eRASS; \cite{Merloni12,Predehl21}), including the five available eRASS observations (see Methods) and the stacked images of the first four eRASS scans (upper limit of 3.98 $\times$ 10$^{-14}$ erg s$^{−1}$ cm$^{−2}$ in the 0.2-2.3 keV energy band). A follow-up campaign with \textit{Swift} started in July 2021 
and revealed that only after February 2024 soft X-ray emission with variations on timescales of days was detected (Extended Data Figure~\ref{fig:ztf}). A \textit{Chandra} Director's Discretionary Time (DDT) observation confirmed the nuclear origin of the emission (ATel \#16576, \cite{Hernandez24}, see Methods). 

\noindent The \textit{NICER} X-ray telescope began a high-cadence monitoring program on May 19th, 2024, with a snapshot roughly every ISS orbit (16x/day, with a minimum separation between consecutive snapshots of 173 seconds and a maximum separation of 9.75 days). The impressive cadence revealed an X-ray light curve consistent with those observed in quasi-periodic eruptions (QPEs; \cite{Miniutti19,Giustini20,Arcodia21,Chakraborty21,Quintin23,Arcodia24b,Nicholl24}). The QPEs in Ansky exhibit burst durations of $\sim$1.5 days, with a $\sim$4.5 day peak-to-peak recurrence time (Methods). We triggered three 30~ks \textit{XMM-Newton} DDT observations, which were taken on June 27th/29th and July 1st, 2024. The \textit{XMM-Newton} observations detected the quiescent accretion disk emission between the QPEs, revealing an 0.3-2 keV flux of $\sim$2 $\times$ 10$^{-14}$ erg s$^{−1}$ cm$^{−2}$ corresponding to a luminosity of 2$\times$10$^{40}$ erg s$^{-1}$. With combined data from \textit{Swift}, \textit{NICER}, and \textit{XMM-Newton}, we detected twelve near-consecutive flares in a 60-day monitoring campaign (Fig.~\ref{fig:lc}). The light curve shows a peculiar super-periodic pattern of five peaks plus a longer gap of $\sim$6.5 days, with hints of decreasing amplitudes within each ``cascade''. We note that during ``cascade 2'' we expected an additional fifth peak (dashed line in Fig.~\ref{fig:lc}) that was missed because no observations were performed during those dates.

\begin{figure}[t!]
 \centering
 \includegraphics[width=\linewidth]{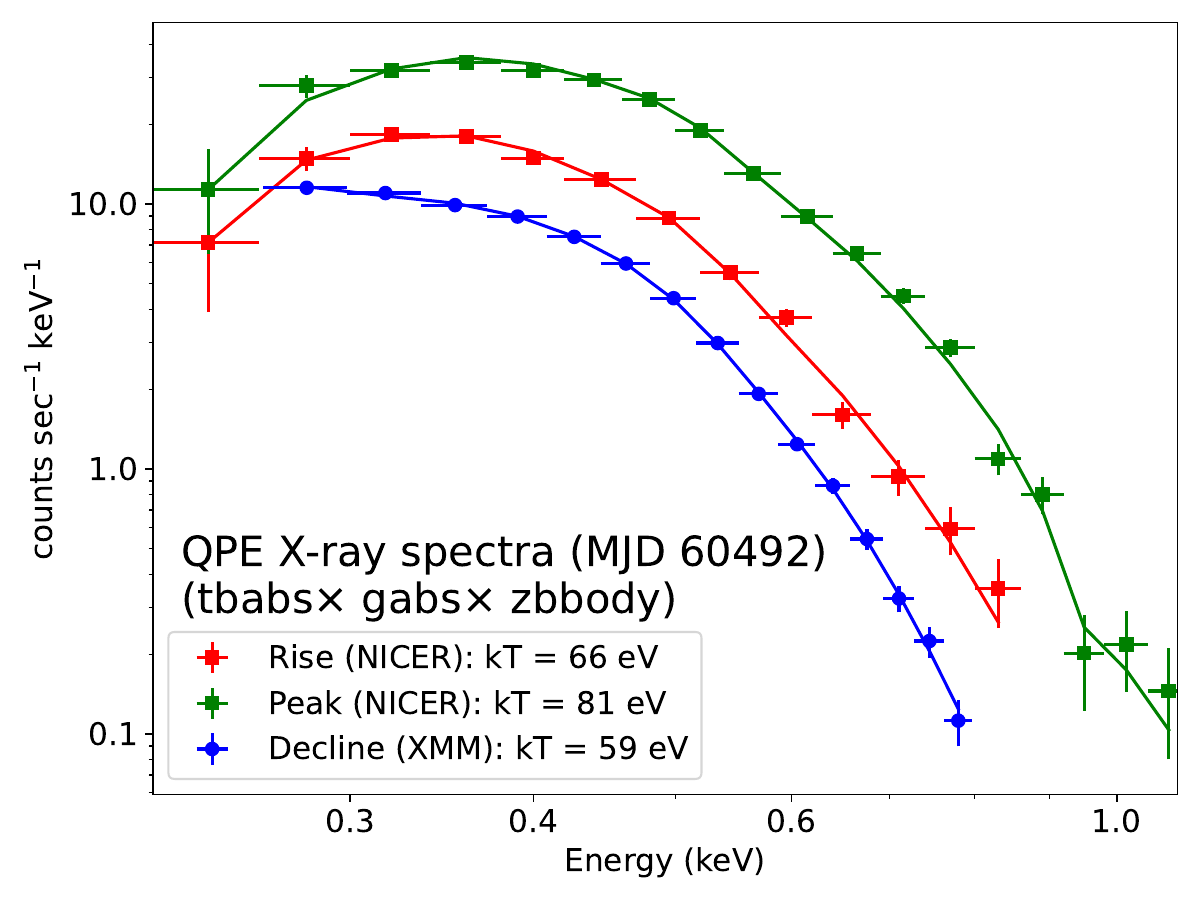}
 \caption{Background-subtracted X-ray spectra of the rise/peak/decline of the QPE on MJD 60492 with \textit{NICER} and \textit{XMM-Newton} data. All phases of the QPE's spectral evolution are well-described by a blackbody with a gaussian absorption component.}
\label{fig:spec}
\end{figure}

\begin{figure}[t!]
 \centering
 \includegraphics[width=\linewidth]{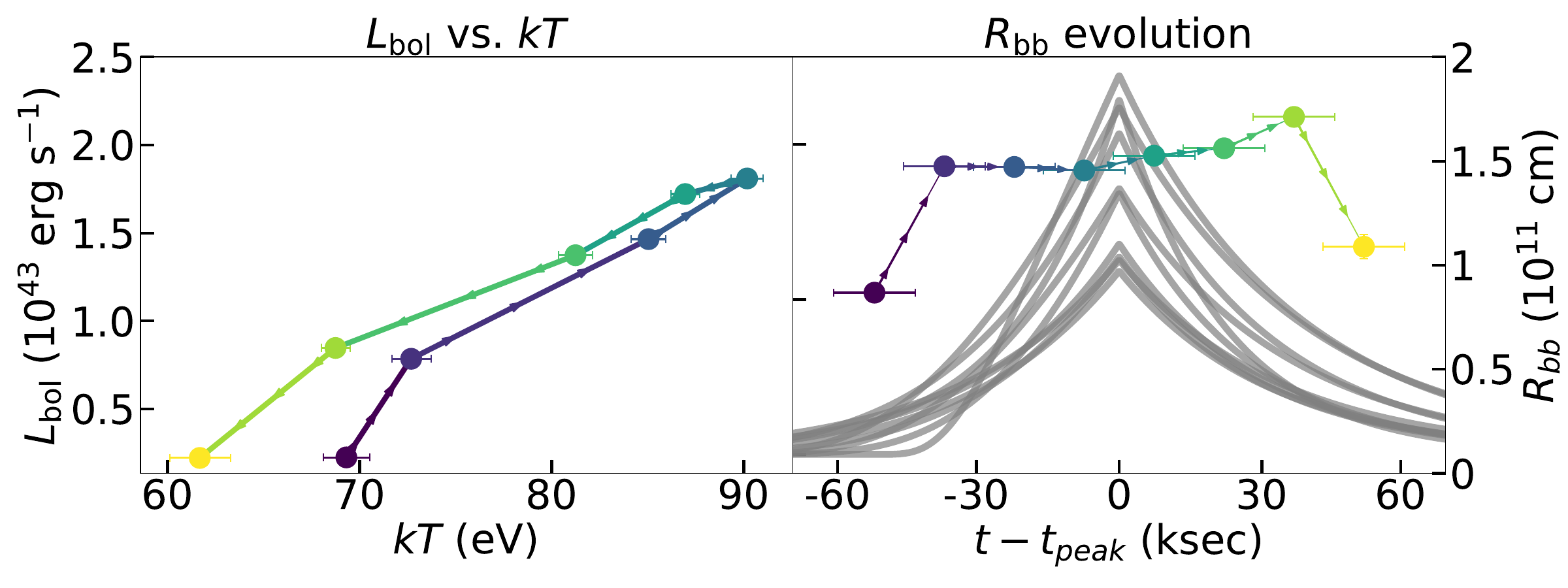}
 \caption{Averaged spectral evolution of the QPEs in Ansky for all well-sampled bursts observed by \nicer. The color scheme indicates the time relative to peak (dark/light at early/late times). Similarly to other sources, the QPEs undergo hysteresis in the $L$-$kT$ plane (left panel), which for a blackbody-like spectrum can be interpreted as an expanding emission region $R_{\mathrm{bb}}$ (right panel).}
\label{fig:hyst}
\end{figure}

\noindent The X-ray spectrum of the QPE is super-soft, being well-described by a black body of varying temperature ($kT\sim 50-100$ eV; Fig.~\ref{fig:spec}; Methods). The spectral evolution throughout the bursts shows the typical ``hotter-when-brighter'' behavior seen in other QPEs, with hints of a counter-clockwise hysteresis pattern in the $L-kT$ plane (Fig.~\ref{fig:hyst}). This is a common feature of other QPEs \cite{Miniutti23a,Arcodia22,Chakraborty24,Giustini24}, and can be modelled with a blackbody emission radius starting at $R_{bb}\sim R_\odot$ growing by a factor of 2--3 over the course of an eruption. An additional absorption component is required around 0.7--1.2 keV, which is sometimes seen in TDEs \cite{Kara18,Masterson22,Yao24} but never before in QPEs. The absorption component shows significant changes over just 10~ks (Methods). For comparison, the only other QPE showing absorption in addition to the continuum is GSN 069. However, that feature does not show such rapid variations, and is probably not directly related to the QPEs themselves but to a distant absorber at $10^{3-4} R_g$ \cite{Miniutti23a,Kosec24}.

\begin{figure}[t!]
 \centering
 \includegraphics[width=\linewidth]{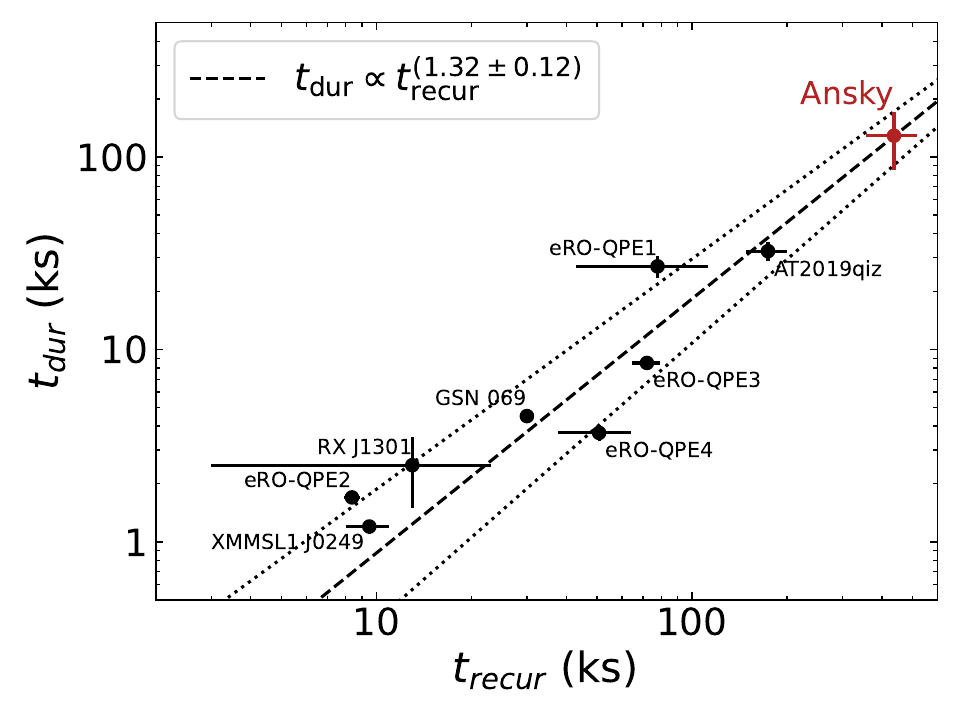}
 \caption{Burst duration, $t_{\mathrm{dur}}$, vs recurrence time, $t_{\mathrm{recur}}$. The best-fit for all known QPEs is a power-law $t_{\mathrm{dur}} \propto t_{\mathrm{recur}}^{1.32}$. Ansky extends the scaling relation by factors of 2.5 and 4, respectively.}
\label{fig:timescales}
\end{figure}

\noindent Ansky extends the maximum timescales of duration and recurrence of known QPEs by a factor $>2.5$ \cite{Arcodia24b,Nicholl24}, inhabiting an intermediate timescale between the QPEs and the diverse, rapidly-growing population of longer-period repeating transients from galactic nuclei, e.g. ASASSN-14ko \cite{Payne21}, eRASSt J0546-20 \cite{Liu23}, and Swift J0230 \cite{Evans23,Guolo24}. In Fig.~\ref{fig:timescales}, we show the burst durations and peak-to-peak recurrence times of all QPEs from the literature alongside Ansky (the QPE candidate AT\,2019vcb was not included because its recurrence time is uncertain \cite{Quintin23,Bykov24}). Fitting the observed data points with a power-law gives $t_{\rm {dur}} \propto t_{\rm {recur}}^{1.32}$. Caveats in this relation might be related to the fact that different QPE sources have durations estimated in somewhat inhomogeneous ways, and the duration also depends
on the effective X-ray band used, which is softer for \textit{XMM-Newton} than for \textit{NICER/Swift} and the duration is known to be energy-dependent \citep{Miniutti19}.
Ansky shows a similar peak luminosity as other QPEs, although the 500x change in amplitude (i.e., the ratio of the peak to quiescent luminosity) is a factor $\sim$10-20x larger than other QPEs. With comparable luminosity, yet much longer timescales, the average integrated energy output per burst of Ansky is $(9.7 \pm {3.7})\times 10^{47}$ ergs -- about 10x higher than other QPEs. This is a useful constraint in constructing theoretical models: QPEs are not characterized by a constant energy which is emitted over a range of timescales. Rather, longer-duration QPEs seem to have larger intrinsic energy budgets altogether.

Observational evidence has accumulated over the past five years associating QPEs with the aftermath of TDEs. Four QPEs are known which occurred following a long-term $t^{-5/3}$ decay in flux \cite{Miniutti19,Chakraborty21,Quintin23,Nicholl24,Bykov24}. The QPE-hosting galaxy GSN 069 shows an abnormally high C/N ratio as seen in TDEs due to CNO processing in the disrupted stellar envelope \cite{Sheng21,Kosec24,Miller23}, and a compact nuclear [O III]-emitting region of $<$35 pc suggesting an accretion system less than 100 yrs old \cite{Patra24}. Long-term monitoring of eRO-QPE1 found a 90x decay in QPE luminosity suggestive of a long-term decline \cite{Chakraborty24,Pasham24}; a similar trend was also seen in eRO-QPE3 \cite{Arcodia24b}. QPEs and TDEs are significantly overrepresented in low-mass post-starburst/quiescent Balmer-strong galaxies, which comprise $\lesssim 0.2\%$ and $2\%$, respectively, of galaxies in the local universe, but are a third of known QPE and 40\% of TDE hosts \cite{Wevers22,Wevers24}; this extreme statistical coincidence suggests a common link to the gas-rich environments of recently faded galactic nuclei.

QPEs are now most commonly modelled as the Extreme Mass-Ratio Inspiral (EMRI) of a stellar-mass object around a supermassive black hole (SMBH), whereby the X-ray emission is generated from shocks between the secondary and the accretion disk of the SMBH \cite{Xian21,Lu23,Franchini23,Linial23,Tagawa23,Zhou24,Arcodia24c}. In this scenario, the nuclei of TDE hosts may generally result in more frequent capture of stellar orbits by the SMBH, conveniently explaining the EMRI companion. The compact disks formed by TDEs should take $\sim$years to viscously spread from their initial formation radius to the EMRI orbital radii, giving a convenient explanation for the several-year delays between the initial transient and the QPEs beginning in GSN 069, and AT2019qiz. Observational estimates of the volumetric rates of QPEs find they are a factor $\sim 10\times$ intrinsically rarer than TDEs \cite{Arcodia24a}, which aligns reasonably well with theoretical expectations for the occurrence fraction of EMRIs at the appropriate orbits \cite{Linial23,Linial24a}. Ansky can also be explained in the EMRI scenario (Methods). The longer duration and recurrence time-scales imply that the EMRI orbit is wider than the other QPEs described in the literature.  Moreover, the repeating pattern of five flares followed by an interruption potentially indicates that the disk has a precessing, eccentric cavity (Methods).
Indeed, super-periodic modulations have been observed before in QPEs \cite{Chakraborty24}, and have been associated with timescales corresponding to disk precession \cite{Franchini23}.

Ansky carries interesting implications for the emerging QPE-TDE connection. It is likely the first time we have witnessed QPEs emerge in real time (see Methods), following an optical transient that is not a ``typical'' TDE, in the sense that Ansky shows no broad H/He lines and shows a power-law decay that is much flatter than the expected decay of TDEs with an index $\sim-5/3$. (\cite{Sanchez24}, Methods). The host galaxy shows the emergence of narrow high-ionization [O\,{\sc iii}] lines; at the same time, the X-ray spectra show no signs of obscuration, making the galaxy similar to a ``bare Seyfert II''. Notably, the original QPE host GSN 069 is also a bare Seyfert II \citep{miniutti2013} galaxy which showed a smooth decline in its X-ray flux before the emergence of QPEs, though the initial emergence of the narrow lines was not constrained.
The optical and X-ray properties of the host galaxies of RX J1301.9+2747 and eRO-QPE2 would also place them as bare Seyfert IIs \citep{Giustini20, Arcodia21}.
Moreover, Ansky has shown low amplitude optical/UV variability since 2020, with increasing amplitude at higher energies (most extreme in the UVW2 band; Fig.~\ref{fig:lc}, Methods).
UV variations were already reported by \cite{Sanchez24} when comparing GALEX data from 2004 and \textit{Swift} data from 2021, which could be related to disk emission. 
The optical/UV flux does not appear to be directly correlated with the X-ray QPEs, though higher-cadence multiwavelength monitoring is underway to determine whether any link between them exists.
The significance of the UV variability, which is still marginal in the existing data, will be studied in follow-up work, and it carries implications for models which predict the QPE emission may extend to low-energy tails \cite{Linial24b}.

These optical/UV variability and spectroscopic properties point towards a newborn AGN-like accretion flow. One interesting possibility is that, with a new reservoir of low angular-momentum gas to accrete, the nucleus of SDSS1335+0728 is now susceptible to a large increase in the rate of stellar capture and disruptions, which has been theorized as a hallmark signature of ``turn-on AGN'' \cite{Starfall22,Ryu24}. 
In such a scenario, the AGN-TDE loss cone would be filled with previous unperturbed stars, a fraction of which would be at the right orbital energies and angular momenta to produce a long-lived EMRI thought to be responsible for the quasi-periodic flares \cite{Linial23}. In this regard, newborn AGN may be an ideal environment to produce QPEs, as they should have rich potential for dynamical interactions between stars in the nuclear cluster plummeting toward the central SMBH.

The QPEs in Ansky show some of the most extreme features that have been observed within this population of objects, including the largest QPE flux and amplitude, longest time scales, and highest energy output,  where the energy budget is $\sim10^{48}$ ergs per burst. Temporal variations are also the most unusual, including a peculiar super-periodic pattern.
On the other hand, while a transient optical event occurred in 2019, it does not correspond to a ``standard" TDE. Instead, the galaxy exhibits signs of an activating AGN or possibly of a new class of SMBH transient event.
In this scenario, a new accretion disk was recently formed, which gets impacted periodically by stellar-mass objects, as proposed for other QPEs. The interruptions every 25 days remain an important challenge for this model, and in the Methods we speculate that they might be explained by disk precession if the latter has an eccentric inner cavity, and/or by a body passing through two different surfaces. 
Continued follow-up of the eruptions in this source will make Ansky one of the finest laboratories for precise tests of QPE models, including probes of the radiation mechanisms and orbital dynamics relevant for EMRI models. Finally, this source represents clear evidence that the QPE-TDE connection is only a subset of a larger trend, whereby QPEs appear not only after ``standard'' TDEs showing broad Balmer/He lines and $t^{-5/3}$ power-law decays, but are more generally associated with newly-formed accretion flows on to massive black holes.

\vspace{0.5cm}

\noindent {\Large \bf References}
\bibliography{main}

\begin{thebibliography}{10}
\expandafter\ifx\csname url\endcsname\relax
  \def\url#1{\texttt{#1}}\fi
\expandafter\ifx\csname urlprefix\endcsname\relax\def\urlprefix{URL }\fi
\providecommand{\bibinfo}[2]{#2}
\providecommand{\eprint}[2][]{\url{#2}}

\bibitem{Sanchez24}
\bibinfo{author}{{S{\'a}nchez-S{\'a}ez}, P.} \emph{et~al.}
\newblock {SDSS1335+0728: The awakening of a {\ensuremath{\sim}}{}10$^{6}$ M$_{{\ensuremath{\odot}}}$ black hole}.
\newblock \emph{\bibinfo{journal}{\aap}} \textbf{\bibinfo{volume}{688}}, \bibinfo{pages}{A157} (\bibinfo{year}{2024}).
\newblock \eprint{2406.11983}.

\bibitem{Martin05}
\bibinfo{author}{{Martin}, D.~C.} \emph{et~al.}
\newblock {The Galaxy Evolution Explorer: A Space Ultraviolet Survey Mission}.
\newblock \emph{\bibinfo{journal}{\apjl}} \textbf{\bibinfo{volume}{619}}, \bibinfo{pages}{L1--L6} (\bibinfo{year}{2005}).
\newblock \eprint{astro-ph/0411302}.

\bibitem{Burrows05}
\bibinfo{author}{{Burrows}, D.~N.} \emph{et~al.}
\newblock {The Swift X-Ray Telescope}.
\newblock \emph{\bibinfo{journal}{\ssr}} \textbf{\bibinfo{volume}{120}}, \bibinfo{pages}{165--195} (\bibinfo{year}{2005}).
\newblock \eprint{astro-ph/0508071}.

\bibitem{Wright10}
\bibinfo{author}{{Wright}, E.~L.} \emph{et~al.}
\newblock {The Wide-field Infrared Survey Explorer (WISE): Mission Description and Initial On-orbit Performance}.
\newblock \emph{\bibinfo{journal}{\aj}} \textbf{\bibinfo{volume}{140}}, \bibinfo{pages}{1868--1881} (\bibinfo{year}{2010}).
\newblock \eprint{1008.0031}.

\bibitem{Mainzer14}
\bibinfo{author}{{Mainzer}, A.} \emph{et~al.}
\newblock {Initial Performance of the NEOWISE Reactivation Mission}.
\newblock \emph{\bibinfo{journal}{\apj}} \textbf{\bibinfo{volume}{792}}, \bibinfo{pages}{30} (\bibinfo{year}{2014}).
\newblock \eprint{1406.6025}.

\bibitem{vanVelzen20}
\bibinfo{author}{{van Velzen}, S.}, \bibinfo{author}{{Holoien}, T. W.~S.}, \bibinfo{author}{{Onori}, F.}, \bibinfo{author}{{Hung}, T.} \& \bibinfo{author}{{Arcavi}, I.}
\newblock {Optical-Ultraviolet Tidal Disruption Events}.
\newblock \emph{\bibinfo{journal}{\ssr}} \textbf{\bibinfo{volume}{216}}, \bibinfo{pages}{124} (\bibinfo{year}{2020}).
\newblock \eprint{2008.05461}.

\bibitem{Reines15}
\bibinfo{author}{{Reines}, A.~E.} \& \bibinfo{author}{{Volonteri}, M.}
\newblock {Relations between Central Black Hole Mass and Total Galaxy Stellar Mass in the Local Universe}.
\newblock \emph{\bibinfo{journal}{\apj}} \textbf{\bibinfo{volume}{813}}, \bibinfo{pages}{82} (\bibinfo{year}{2015}).
\newblock \eprint{1508.06274}.

\bibitem{Merloni12}
\bibinfo{author}{{Merloni}, A.} \emph{et~al.}
\newblock {eROSITA Science Book: Mapping the Structure of the Energetic Universe}.
\newblock \emph{\bibinfo{journal}{arXiv e-prints}} \bibinfo{pages}{arXiv:1209.3114} (\bibinfo{year}{2012}).
\newblock \eprint{1209.3114}.

\bibitem{Predehl21}
\bibinfo{author}{{Predehl}, P.} \emph{et~al.}
\newblock {The eROSITA X-ray telescope on SRG}.
\newblock \emph{\bibinfo{journal}{\aap}} \textbf{\bibinfo{volume}{647}}, \bibinfo{pages}{A1} (\bibinfo{year}{2021}).
\newblock \eprint{2010.03477}.

\bibitem{Hernandez24}
\bibinfo{author}{{Hernandez-Garcia}, L.} \emph{et~al.}
\newblock {Swift and Chandra X-ray detections in the nucleus of SDSS J133519.91+072807.4}.
\newblock \emph{\bibinfo{journal}{The Astronomer's Telegram}} \textbf{\bibinfo{volume}{16576}}, \bibinfo{pages}{1} (\bibinfo{year}{2024}).

\bibitem{Miniutti19}
\bibinfo{author}{{Miniutti}, G.} \emph{et~al.}
\newblock {Nine-hour X-ray quasi-periodic eruptions from a low-mass black hole galactic nucleus}.
\newblock \emph{\bibinfo{journal}{\nat}} \textbf{\bibinfo{volume}{573}}, \bibinfo{pages}{381--384} (\bibinfo{year}{2019}).
\newblock \eprint{1909.04693}.

\bibitem{Giustini20}
\bibinfo{author}{{Giustini}, M.}, \bibinfo{author}{{Miniutti}, G.} \& \bibinfo{author}{{Saxton}, R.~D.}
\newblock {X-ray quasi-periodic eruptions from the galactic nucleus of RX J1301.9+2747}.
\newblock \emph{\bibinfo{journal}{\aap}} \textbf{\bibinfo{volume}{636}}, \bibinfo{pages}{L2} (\bibinfo{year}{2020}).
\newblock \eprint{2002.08967}.

\bibitem{Arcodia21}
\bibinfo{author}{{Arcodia}, R.} \emph{et~al.}
\newblock {X-ray quasi-periodic eruptions from two previously quiescent galaxies}.
\newblock \emph{\bibinfo{journal}{\nat}} \textbf{\bibinfo{volume}{592}}, \bibinfo{pages}{704--707} (\bibinfo{year}{2021}).
\newblock \eprint{2104.13388}.

\bibitem{Chakraborty21}
\bibinfo{author}{{Chakraborty}, J.} \emph{et~al.}
\newblock {Possible X-Ray Quasi-periodic Eruptions in a Tidal Disruption Event Candidate}.
\newblock \emph{\bibinfo{journal}{\apjl}} \textbf{\bibinfo{volume}{921}}, \bibinfo{pages}{L40} (\bibinfo{year}{2021}).
\newblock \eprint{2110.10786}.

\bibitem{Quintin23}
\bibinfo{author}{{Quintin}, E.} \emph{et~al.}
\newblock {Tormund's return: Hints of quasi-periodic eruption features from a recent optical tidal disruption event}.
\newblock \emph{\bibinfo{journal}{\aap}} \textbf{\bibinfo{volume}{675}}, \bibinfo{pages}{A152} (\bibinfo{year}{2023}).
\newblock \eprint{2306.00438}.

\bibitem{Arcodia24b}
\bibinfo{author}{{Arcodia}, R.} \emph{et~al.}
\newblock {The more the merrier: SRG/eROSITA discovers two further galaxies showing X-ray quasi-periodic eruptions}.
\newblock \emph{\bibinfo{journal}{\aap}} \textbf{\bibinfo{volume}{684}}, \bibinfo{pages}{A64} (\bibinfo{year}{2024}).
\newblock \eprint{2401.17275}.

\bibitem{Nicholl24}
\bibinfo{author}{{Nicholl}, M.} \emph{et~al.}
\newblock {Quasi-periodic X-ray eruptions years after a nearby tidal disruption event}.
\newblock \emph{\bibinfo{journal}{arXiv e-prints}} \bibinfo{pages}{arXiv:2409.02181} (\bibinfo{year}{2024}).
\newblock \eprint{2409.02181}.

\bibitem{Miniutti23a}
\bibinfo{author}{{Miniutti}, G.} \emph{et~al.}
\newblock {Repeating tidal disruptions in GSN 069: Long-term evolution and constraints on quasi-periodic eruptions' models}.
\newblock \emph{\bibinfo{journal}{\aap}} \textbf{\bibinfo{volume}{670}}, \bibinfo{pages}{A93} (\bibinfo{year}{2023}).
\newblock \eprint{2207.07511}.

\bibitem{Arcodia22}
\bibinfo{author}{{Arcodia}, R.} \emph{et~al.}
\newblock {The complex time and energy evolution of quasi-periodic eruptions in eRO-QPE1}.
\newblock \emph{\bibinfo{journal}{\aap}} \textbf{\bibinfo{volume}{662}}, \bibinfo{pages}{A49} (\bibinfo{year}{2022}).
\newblock \eprint{2203.11939}.

\bibitem{Chakraborty24}
\bibinfo{author}{{Chakraborty}, J.} \emph{et~al.}
\newblock {Testing EMRI Models for Quasi-periodic Eruptions with 3.5 yr of Monitoring eRO-QPE1}.
\newblock \emph{\bibinfo{journal}{\apj}} \textbf{\bibinfo{volume}{965}}, \bibinfo{pages}{12} (\bibinfo{year}{2024}).
\newblock \eprint{2402.08722}.

\bibitem{Giustini24}
\bibinfo{author}{{Giustini}, M.} \emph{et~al.}
\newblock {Fragments of harmony amid apparent chaos: a closer look at the X-ray quasi-periodic eruptions of the galaxy RX J1301.9+2747}.
\newblock \emph{\bibinfo{journal}{arXiv e-prints}} \bibinfo{pages}{arXiv:2409.01938} (\bibinfo{year}{2024}).
\newblock \eprint{2409.01938}.

\bibitem{Kara18}
\bibinfo{author}{{Kara}, E.}, \bibinfo{author}{{Dai}, L.}, \bibinfo{author}{{Reynolds}, C.~S.} \& \bibinfo{author}{{Kallman}, T.}
\newblock {Ultrafast outflow in tidal disruption event ASASSN-14li}.
\newblock \emph{\bibinfo{journal}{\mnras}} \textbf{\bibinfo{volume}{474}}, \bibinfo{pages}{3593--3598} (\bibinfo{year}{2018}).
\newblock \eprint{1711.06090}.

\bibitem{Masterson22}
\bibinfo{author}{{Masterson}, M.} \emph{et~al.}
\newblock {Evolution of a Relativistic Outflow and X-Ray Corona in the Extreme Changing-look AGN 1ES 1927+654}.
\newblock \emph{\bibinfo{journal}{\apj}} \textbf{\bibinfo{volume}{934}}, \bibinfo{pages}{35} (\bibinfo{year}{2022}).
\newblock \eprint{2206.05140}.

\bibitem{Yao24}
\bibinfo{author}{{Yao}, Y.} \emph{et~al.}
\newblock {Sub-relativistic Outflow and Hours-Timescale Large-amplitude X-ray Dips during Super-Eddington Accretion onto a Low-mass Massive Black Hole in the Tidal Disruption Event AT2022lri}.
\newblock \emph{\bibinfo{journal}{arXiv e-prints}} \bibinfo{pages}{arXiv:2405.11343} (\bibinfo{year}{2024}).
\newblock \eprint{2405.11343}.

\bibitem{Kosec24}
\bibinfo{author}{{Kosec}, P.} \emph{et~al.}
\newblock {Detection of a Highly Ionized Outflow in the Quasi-periodically Erupting Source GSN 069}.
\newblock \emph{\bibinfo{journal}{arXiv e-prints}} \bibinfo{pages}{arXiv:2406.17105} (\bibinfo{year}{2024}).
\newblock \eprint{2406.17105}.

\bibitem{Payne21}
\bibinfo{author}{{Payne}, A.~V.} \emph{et~al.}
\newblock {ASASSN-14ko is a Periodic Nuclear Transient in ESO 253-G003}.
\newblock \emph{\bibinfo{journal}{\apj}} \textbf{\bibinfo{volume}{910}}, \bibinfo{pages}{125} (\bibinfo{year}{2021}).
\newblock \eprint{2009.03321}.

\bibitem{Liu23}
\bibinfo{author}{{Liu}, Z.} \emph{et~al.}
\newblock {Deciphering the extreme X-ray variability of the nuclear transient eRASSt J045650.3{\ensuremath{-}}203750. A likely repeating partial tidal disruption event}.
\newblock \emph{\bibinfo{journal}{\aap}} \textbf{\bibinfo{volume}{669}}, \bibinfo{pages}{A75} (\bibinfo{year}{2023}).
\newblock \eprint{2208.12452}.

\bibitem{Evans23}
\bibinfo{author}{{Evans}, P.~A.} \emph{et~al.}
\newblock {Monthly quasi-periodic eruptions from repeated stellar disruption by a massive black hole}.
\newblock \emph{\bibinfo{journal}{Nature Astronomy}} \textbf{\bibinfo{volume}{7}}, \bibinfo{pages}{1368--1375} (\bibinfo{year}{2023}).
\newblock \eprint{2309.02500}.

\bibitem{Guolo24}
\bibinfo{author}{{Guolo}, M.} \emph{et~al.}
\newblock {X-ray eruptions every 22 days from the nucleus of a nearby galaxy}.
\newblock \emph{\bibinfo{journal}{Nature Astronomy}} \textbf{\bibinfo{volume}{8}}, \bibinfo{pages}{347--358} (\bibinfo{year}{2024}).
\newblock \eprint{2309.03011}.

\bibitem{Bykov24}
\bibinfo{author}{{Bykov}, S.}, \bibinfo{author}{{Gilfanov}, M.}, \bibinfo{author}{{Sunyaev}, R.} \& \bibinfo{author}{{Medvedev}, P.}
\newblock {Further evidence of Quasiperiodic Eruptions in a tidal disruption event AT2019vcb by SRG/eROSITA}.
\newblock \emph{\bibinfo{journal}{arXiv e-prints}} \bibinfo{pages}{arXiv:2409.16908} (\bibinfo{year}{2024}).
\newblock \eprint{2409.16908}.

\bibitem{Sheng21}
\bibinfo{author}{{Sheng}, Z.} \emph{et~al.}
\newblock {Evidence of a Tidal-disruption Event in GSN 069 from the Abnormal Carbon and Nitrogen Abundance Ratio}.
\newblock \emph{\bibinfo{journal}{\apjl}} \textbf{\bibinfo{volume}{920}}, \bibinfo{pages}{L25} (\bibinfo{year}{2021}).
\newblock \eprint{2109.01683}.

\bibitem{Miller23}
\bibinfo{author}{{Miller}, J.~M.} \emph{et~al.}
\newblock {Evidence of a Massive Stellar Disruption in the X-Ray Spectrum of ASASSN-14li}.
\newblock \emph{\bibinfo{journal}{\apjl}} \textbf{\bibinfo{volume}{953}}, \bibinfo{pages}{L23} (\bibinfo{year}{2023}).
\newblock \eprint{2308.10964}.

\bibitem{Patra24}
\bibinfo{author}{{Patra}, K.~C.} \emph{et~al.}
\newblock {Constraints on the narrow-line region of the X-ray quasi-periodic eruption source GSN 069}.
\newblock \emph{\bibinfo{journal}{\mnras}} \textbf{\bibinfo{volume}{530}}, \bibinfo{pages}{5120--5130} (\bibinfo{year}{2024}).
\newblock \eprint{2310.05574}.

\bibitem{Pasham24}
\bibinfo{author}{{Pasham}, D.~R.} \emph{et~al.}
\newblock {Alive but Barely Kicking: News from 3+ yr of Swift and XMM-Newton X-Ray Monitoring of Quasiperiodic Eruptions from eRO-QPE1}.
\newblock \emph{\bibinfo{journal}{\apjl}} \textbf{\bibinfo{volume}{963}}, \bibinfo{pages}{L47} (\bibinfo{year}{2024}).
\newblock \eprint{2402.09690}.

\bibitem{Wevers22}
\bibinfo{author}{{Wevers}, T.}, \bibinfo{author}{{Pasham}, D.~R.}, \bibinfo{author}{{Jalan}, P.}, \bibinfo{author}{{Rakshit}, S.} \& \bibinfo{author}{{Arcodia}, R.}
\newblock {Host galaxy properties of quasi-periodically erupting X-ray sources}.
\newblock \emph{\bibinfo{journal}{\aap}} \textbf{\bibinfo{volume}{659}}, \bibinfo{pages}{L2} (\bibinfo{year}{2022}).
\newblock \eprint{2201.11751}.

\bibitem{Wevers24}
\bibinfo{author}{{Wevers}, T.} \emph{et~al.}
\newblock {X-Ray Quasi-periodic Eruptions and Tidal Disruption Events Prefer Similar Host Galaxies}.
\newblock \emph{\bibinfo{journal}{\apjl}} \textbf{\bibinfo{volume}{970}}, \bibinfo{pages}{L23} (\bibinfo{year}{2024}).
\newblock \eprint{2406.02678}.

\bibitem{Xian21}
\bibinfo{author}{{Xian}, J.}, \bibinfo{author}{{Zhang}, F.}, \bibinfo{author}{{Dou}, L.}, \bibinfo{author}{{He}, J.} \& \bibinfo{author}{{Shu}, X.}
\newblock {X-Ray Quasi-periodic Eruptions Driven by Star-Disk Collisions: Application to GSN069 and Probing the Spin of Massive Black Holes}.
\newblock \emph{\bibinfo{journal}{\apjl}} \textbf{\bibinfo{volume}{921}}, \bibinfo{pages}{L32} (\bibinfo{year}{2021}).
\newblock \eprint{2110.10855}.

\bibitem{Lu23}
\bibinfo{author}{{Lu}, W.} \& \bibinfo{author}{{Quataert}, E.}
\newblock {Quasi-periodic eruptions from mildly eccentric unstable mass transfer in galactic nuclei}.
\newblock \emph{\bibinfo{journal}{\mnras}} \textbf{\bibinfo{volume}{524}}, \bibinfo{pages}{6247--6266} (\bibinfo{year}{2023}).
\newblock \eprint{2210.08023}.

\bibitem{Franchini23}
\bibinfo{author}{{Franchini}, A.} \emph{et~al.}
\newblock {Quasi-periodic eruptions from impacts between the secondary and a rigidly precessing accretion disc in an extreme mass-ratio inspiral system}.
\newblock \emph{\bibinfo{journal}{\aap}} \textbf{\bibinfo{volume}{675}}, \bibinfo{pages}{A100} (\bibinfo{year}{2023}).
\newblock \eprint{2304.00775}.

\bibitem{Linial23}
\bibinfo{author}{{Linial}, I.} \& \bibinfo{author}{{Metzger}, B.~D.}
\newblock {EMRI + TDE = QPE: Periodic X-Ray Flares from Star-Disk Collisions in Galactic Nuclei}.
\newblock \emph{\bibinfo{journal}{\apj}} \textbf{\bibinfo{volume}{957}}, \bibinfo{pages}{34} (\bibinfo{year}{2023}).
\newblock \eprint{2303.16231}.

\bibitem{Tagawa23}
\bibinfo{author}{{Tagawa}, H.} \& \bibinfo{author}{{Haiman}, Z.}
\newblock {Flares from stars crossing active galactic nucleus discs on low-inclination orbits}.
\newblock \emph{\bibinfo{journal}{\mnras}} \textbf{\bibinfo{volume}{526}}, \bibinfo{pages}{69--79} (\bibinfo{year}{2023}).
\newblock \eprint{2304.03670}.

\bibitem{Zhou24}
\bibinfo{author}{{Zhou}, C.}, \bibinfo{author}{{Huang}, L.}, \bibinfo{author}{{Guo}, K.}, \bibinfo{author}{{Li}, Y.-P.} \& \bibinfo{author}{{Pan}, Z.}
\newblock {Probing orbits of stellar mass objects deep in galactic nuclei with quasiperiodic eruptions}.
\newblock \emph{\bibinfo{journal}{\prd}} \textbf{\bibinfo{volume}{109}}, \bibinfo{pages}{103031} (\bibinfo{year}{2024}).
\newblock \eprint{2401.11190}.

\bibitem{Arcodia24c}
\bibinfo{author}{{Arcodia}, R.} \emph{et~al.}
\newblock {Ticking away: the long-term X-ray timing and spectral evolution of eRO-QPE2}.
\newblock \emph{\bibinfo{journal}{arXiv e-prints}} \bibinfo{pages}{arXiv:2406.17020} (\bibinfo{year}{2024}).
\newblock \eprint{2406.17020}.

\bibitem{Arcodia24a}
\bibinfo{author}{{Arcodia}, R.} \emph{et~al.}
\newblock {Cosmic hide and seek: The volumetric rate of X-ray quasi-periodic eruptions}.
\newblock \emph{\bibinfo{journal}{\aap}} \textbf{\bibinfo{volume}{684}}, \bibinfo{pages}{L14} (\bibinfo{year}{2024}).
\newblock \eprint{2403.17059}.

\bibitem{Linial24a}
\bibinfo{author}{{Linial}, I.} \& \bibinfo{author}{{Quataert}, E.}
\newblock {Period evolution of repeating transients in galactic nuclei}.
\newblock \emph{\bibinfo{journal}{\mnras}} \textbf{\bibinfo{volume}{527}}, \bibinfo{pages}{4317--4329} (\bibinfo{year}{2024}).
\newblock \eprint{2309.15849}.

\bibitem{miniutti2013}
\bibinfo{author}{{Miniutti}, G.} \emph{et~al.}
\newblock {A high Eddington-ratio, true Seyfert 2 galaxy candidate: implications for broad-line region models}.
\newblock \emph{\bibinfo{journal}{\mnras}} \textbf{\bibinfo{volume}{433}}, \bibinfo{pages}{1764--1777} (\bibinfo{year}{2013}).
\newblock \eprint{1305.3284}.

\bibitem{Linial24b}
\bibinfo{author}{{Linial}, I.} \& \bibinfo{author}{{Metzger}, B.~D.}
\newblock {Ultraviolet Quasiperiodic Eruptions from Star{\textendash}Disk Collisions in Galactic Nuclei}.
\newblock \emph{\bibinfo{journal}{\apjl}} \textbf{\bibinfo{volume}{963}}, \bibinfo{pages}{L1} (\bibinfo{year}{2024}).
\newblock \eprint{2311.16231}.

\bibitem{Starfall22}
\bibinfo{author}{{McKernan}, B.} \emph{et~al.}
\newblock {Starfall: a heavy rain of stars in 'turning on' AGN}.
\newblock \emph{\bibinfo{journal}{\mnras}} \textbf{\bibinfo{volume}{514}}, \bibinfo{pages}{4102--4110} (\bibinfo{year}{2022}).
\newblock \eprint{2110.03741}.

\bibitem{Ryu24}
\bibinfo{author}{{Ryu}, T.} \emph{et~al.}
\newblock {In-plane tidal disruption of stars in discs of active galactic nuclei}.
\newblock \emph{\bibinfo{journal}{\mnras}} \textbf{\bibinfo{volume}{527}}, \bibinfo{pages}{8103--8117} (\bibinfo{year}{2024}).
\newblock \eprint{2310.00610}.

\bibitem{1992ARA&A..30..575C}
\bibinfo{author}{{Condon}, J.~J.}
\newblock {Radio emission from normal galaxies.}
\newblock \emph{\bibinfo{journal}{\araa}} \textbf{\bibinfo{volume}{30}}, \bibinfo{pages}{575--611} (\bibinfo{year}{1992}).


\expandafter\ifx\csname url\endcsname\relax
  \def\url#1{\texttt{#1}}\fi
\expandafter\ifx\csname urlprefix\endcsname\relax\def\urlprefix{URL }\fi
\providecommand{\bibinfo}[2]{#2}
\providecommand{\eprint}[2][]{\url{#2}}

\bibitem{forster2021}
\bibinfo{author}{{F{\"o}rster}, F.} \emph{et~al.}
\newblock {The Automatic Learning for the Rapid Classification of Events (ALeRCE) Alert Broker}.
\newblock \emph{\bibinfo{journal}{\aj}} \textbf{\bibinfo{volume}{161}}, \bibinfo{pages}{242} (\bibinfo{year}{2021}).
\newblock \eprint{2008.03303}.

\bibitem{Sanchez-Saez21a}
\bibinfo{author}{{S{\'a}nchez-S{\'a}ez}, P.} \emph{et~al.}
\newblock {Alert Classification for the ALeRCE Broker System: The Light Curve Classifier}.
\newblock \emph{\bibinfo{journal}{\aj}} \textbf{\bibinfo{volume}{161}}, \bibinfo{pages}{141} (\bibinfo{year}{2021}).
\newblock \eprint{2008.03311}.

\bibitem{Trakhtenbrot19NatAs}
\bibinfo{author}{{Trakhtenbrot}, B.} \emph{et~al.}
\newblock {A new class of flares from accreting supermassive black holes}.
\newblock \emph{\bibinfo{journal}{Nature Astronomy}} \textbf{\bibinfo{volume}{3}}, \bibinfo{pages}{242--250} (\bibinfo{year}{2019}).
\newblock \eprint{1901.03731}.

\bibitem{Baldwin81}
\bibinfo{author}{{Baldwin}, J.~A.}, \bibinfo{author}{{Phillips}, M.~M.} \& \bibinfo{author}{{Terlevich}, R.}
\newblock {Classification parameters for the emission-line spectra of extragalactic objects.}
\newblock \emph{\bibinfo{journal}{\pasp}} \textbf{\bibinfo{volume}{93}}, \bibinfo{pages}{5--19} (\bibinfo{year}{1981}).

\bibitem{Lyu19}
\bibinfo{author}{{Lyu}, J.}, \bibinfo{author}{{Rieke}, G.~H.} \& \bibinfo{author}{{Smith}, P.~S.}
\newblock {Mid-IR Variability and Dust Reverberation Mapping of Low-z Quasars. I. Data, Methods, and Basic Results}.
\newblock \emph{\bibinfo{journal}{\apj}} \textbf{\bibinfo{volume}{886}}, \bibinfo{pages}{33} (\bibinfo{year}{2019}).
\newblock \eprint{1909.11101}.

\bibitem{Arevalo24}
\bibinfo{author}{{Ar{\'e}valo}, P.} \emph{et~al.}
\newblock {A newborn active galactic nucleus in a star-forming galaxy}.
\newblock \emph{\bibinfo{journal}{\aap}} \textbf{\bibinfo{volume}{683}}, \bibinfo{pages}{L8} (\bibinfo{year}{2024}).
\newblock \eprint{2402.19403}.

\bibitem{Burrows05}
\bibinfo{author}{{Burrows}, D.~N.} \emph{et~al.}
\newblock {The Swift X-Ray Telescope}.
\newblock \emph{\bibinfo{journal}{\ssr}} \textbf{\bibinfo{volume}{120}}, \bibinfo{pages}{165--195} (\bibinfo{year}{2005}).
\newblock \eprint{astro-ph/0508071}.

\bibitem{cash1979}
\bibinfo{author}{{Cash}, W.}
\newblock {Parameter estimation in astronomy through application of the likelihood ratio.}
\newblock \emph{\bibinfo{journal}{\apj}} \textbf{\bibinfo{volume}{228}}, \bibinfo{pages}{939--947} (\bibinfo{year}{1979}).

\bibitem{LSXPSCatalogue2023}
\bibinfo{author}{{Evans}, P.~A.} \emph{et~al.}
\newblock {A real-time transient detector and the living Swift-XRT point source catalogue}.
\newblock \emph{\bibinfo{journal}{\mnras}} \textbf{\bibinfo{volume}{518}}, \bibinfo{pages}{174--184} (\bibinfo{year}{2023}).
\newblock \eprint{2208.14478}.

\bibitem{Weisskopf2000}
\bibinfo{author}{{Weisskopf}, M.~C.}, \bibinfo{author}{{Tananbaum}, H.~D.}, \bibinfo{author}{{Van Speybroeck}, L.~P.} \& \bibinfo{author}{{O'Dell}, S.~L.}
\newblock {Chandra X-ray Observatory (CXO): overview}.
\newblock In \bibinfo{editor}{{Truemper}, J.~E.} \& \bibinfo{editor}{{Aschenbach}, B.} (eds.) \emph{\bibinfo{booktitle}{X-Ray Optics, Instruments, and Missions III}}, vol. \bibinfo{volume}{4012} of \emph{\bibinfo{series}{SPIE Conference Series}}, \bibinfo{pages}{2--16} (\bibinfo{year}{2000}).
\newblock \eprint{astro-ph/0004127}.

\bibitem{Gendreau16}
\bibinfo{author}{{Gendreau}, K.~C.} \emph{et~al.}
\newblock {The Neutron star Interior Composition Explorer (NICER): design and development}.
\newblock In \bibinfo{editor}{{den Herder}, J.-W.~A.}, \bibinfo{editor}{{Takahashi}, T.} \& \bibinfo{editor}{{Bautz}, M.} (eds.) \emph{\bibinfo{booktitle}{Space Telescopes and Instrumentation 2016: Ultraviolet to Gamma Ray}}, vol. \bibinfo{volume}{9905} of \emph{\bibinfo{series}{SPIE Conference Series}}, \bibinfo{pages}{99051H} (\bibinfo{year}{2016}).

\bibitem{Arnaud96}
\bibinfo{author}{{Arnaud}, K.~A.}
\newblock {XSPEC: The First Ten Years}.
\newblock In \bibinfo{editor}{{Jacoby}, G.~H.} \& \bibinfo{editor}{{Barnes}, J.} (eds.) \emph{\bibinfo{booktitle}{Astronomical Data Analysis Software and Systems V}}, vol. \bibinfo{volume}{101} of \emph{\bibinfo{series}{Astronomical Society of the Pacific Conference Series}}, \bibinfo{pages}{17} (\bibinfo{year}{1996}).

\bibitem{Kaastra16}
\bibinfo{author}{{Kaastra}, J.~S.} \& \bibinfo{author}{{Bleeker}, J.~A.~M.}
\newblock {Optimal binning of X-ray spectra and response matrix design}.
\newblock \emph{\bibinfo{journal}{\aap}} \textbf{\bibinfo{volume}{587}}, \bibinfo{pages}{A151} (\bibinfo{year}{2016}).
\newblock \eprint{1601.05309}.

\bibitem{ricci2020}
\bibinfo{author}{{Ricci}, C.} \emph{et~al.}
\newblock {The Destruction and Recreation of the X-Ray Corona in a Changing-look Active Galactic Nucleus}.
\newblock \emph{\bibinfo{journal}{\apjl}} \textbf{\bibinfo{volume}{898}}, \bibinfo{pages}{L1} (\bibinfo{year}{2020}).
\newblock \eprint{2007.07275}.

\bibitem{Chakraborty25}
\bibinfo{author}{{Chakraborty}, J.} \emph{et~al.}
\newblock {Rapidly evolving ionization features in a Quasi-periodic Eruption: spectroscopic evidence for homologous expansion of the emission surface?}
\newblock \emph{\bibinfo{journal}{submitted to ApJ}} .

\bibitem{Tubin24}
\bibinfo{author}{{Tub{\'\i}n-Arenas}, D.} \emph{et~al.}
\newblock {The eROSITA upper limits. Description and access to the data}.
\newblock \emph{\bibinfo{journal}{\aap}} \textbf{\bibinfo{volume}{682}}, \bibinfo{pages}{A35} (\bibinfo{year}{2024}).
\newblock \eprint{2401.17305}.

\bibitem{graham2019}
\bibinfo{author}{{Graham}, M.~J.} \emph{et~al.}
\newblock {The Zwicky Transient Facility: Science Objectives}.
\newblock \emph{\bibinfo{journal}{\pasp}} \textbf{\bibinfo{volume}{131}}, \bibinfo{pages}{078001} (\bibinfo{year}{2019}).
\newblock \eprint{1902.01945}.

\bibitem{bellm2019}
\bibinfo{author}{{Bellm}, E.~C.} \emph{et~al.}
\newblock {The Zwicky Transient Facility: System Overview, Performance, and First Results}.
\newblock \emph{\bibinfo{journal}{\pasp}} \textbf{\bibinfo{volume}{131}}, \bibinfo{pages}{018002} (\bibinfo{year}{2019}).
\newblock \eprint{1902.01932}.

\bibitem{masci2019}
\bibinfo{author}{{Masci}, F.~J.} \emph{et~al.}
\newblock {The Zwicky Transient Facility: Data Processing, Products, and Archive}.
\newblock \emph{\bibinfo{journal}{\pasp}} \textbf{\bibinfo{volume}{131}}, \bibinfo{pages}{018003} (\bibinfo{year}{2019}).
\newblock \eprint{1902.01872}.

\bibitem{masci2023}
\bibinfo{author}{{Masci}, F.~J.} \emph{et~al.}
\newblock {A New Forced Photometry Service for the Zwicky Transient Facility}.
\newblock \emph{\bibinfo{journal}{arXiv e-prints}} \bibinfo{pages}{arXiv:2305.16279} (\bibinfo{year}{2023}).
\newblock \eprint{2305.16279}.

\bibitem{2005SSRv..120...95R}
\bibinfo{author}{{Roming}, P. W.~A.} \emph{et~al.}
\newblock {The Swift Ultra-Violet/Optical Telescope}.
\newblock \emph{\bibinfo{journal}{\ssr}} \textbf{\bibinfo{volume}{120}}, \bibinfo{pages}{95--142} (\bibinfo{year}{2005}).
\newblock \eprint{astro-ph/0507413}.

\bibitem{2023GCNswift}
\bibinfo{author}{{Cenko}, B.}
\newblock {Swift Attitude Control Affecting Some UVOT Images}.
\newblock \emph{\bibinfo{journal}{GRB Coordinates Network}} \textbf{\bibinfo{volume}{34633}}, \bibinfo{pages}{1} (\bibinfo{year}{2023}).

\bibitem{2024GCNswift}
{Swift satellite resumes pointed science observations}.
\newblock \emph{\bibinfo{journal}{GRB Coordinates Network}} \textbf{\bibinfo{volume}{36033}}, \bibinfo{pages}{1} (\bibinfo{year}{2024}).

\bibitem{Rees88}
\bibinfo{author}{{Rees}, M.~J.}
\newblock {Tidal disruption of stars by black holes of {}10$^{6}$-{}10$^{8}$ solar masses in nearby galaxies}.
\newblock \emph{\bibinfo{journal}{\nat}} \textbf{\bibinfo{volume}{333}}, \bibinfo{pages}{523--528} (\bibinfo{year}{1988}).

\bibitem{Phinney89}
\bibinfo{author}{{Phinney}, E.~S.}
\newblock {Manifestations of a Massive Black Hole in the Galactic Center}.
\newblock In \bibinfo{editor}{{Morris}, M.} (ed.) \emph{\bibinfo{booktitle}{The Center of the Galaxy}}, vol. \bibinfo{volume}{136}, \bibinfo{pages}{543} (\bibinfo{year}{1989}).

\bibitem{Hammerstein23}
\bibinfo{author}{{Hammerstein}, E.} \emph{et~al.}
\newblock {The Final Season Reimagined: 30 Tidal Disruption Events from the ZTF-I Survey}.
\newblock \emph{\bibinfo{journal}{\apj}} \textbf{\bibinfo{volume}{942}}, \bibinfo{pages}{9} (\bibinfo{year}{2023}).
\newblock \eprint{2203.01461}.

\bibitem{vaughan2003}
\bibinfo{author}{{Vaughan}, S.}, \bibinfo{author}{{Edelson}, R.}, \bibinfo{author}{{Warwick}, R.~S.} \& \bibinfo{author}{{Uttley}, P.}
\newblock {On characterizing the variability properties of X-ray light curves from active galaxies}.
\newblock \emph{\bibinfo{journal}{\mnras}} \textbf{\bibinfo{volume}{345}}, \bibinfo{pages}{1271--1284} (\bibinfo{year}{2003}).
\newblock \eprint{astro-ph/0307420}.

\bibitem{CASA2022}
\bibinfo{author}{{CASA Team}} \emph{et~al.}
\newblock {CASA, the Common Astronomy Software Applications for Radio Astronomy}.
\newblock \emph{\bibinfo{journal}{\pasp}} \textbf{\bibinfo{volume}{134}}, \bibinfo{pages}{114501} (\bibinfo{year}{2022}).
\newblock \eprint{2210.02276}.

\bibitem{2017A&A...598A..78I}
\bibinfo{author}{{Intema}, H.~T.}, \bibinfo{author}{{Jagannathan}, P.}, \bibinfo{author}{{Mooley}, K.~P.} \& \bibinfo{author}{{Frail}, D.~A.}
\newblock {The GMRT 150 MHz all-sky radio survey. First alternative data release TGSS ADR1}.
\newblock \emph{\bibinfo{journal}{\aap}} \textbf{\bibinfo{volume}{598}}, \bibinfo{pages}{A78} (\bibinfo{year}{2017}).
\newblock \eprint{1603.04368}.

\bibitem{2020PASA...37...48M}
\bibinfo{author}{{McConnell}, D.} \emph{et~al.}
\newblock {The Rapid ASKAP Continuum Survey I: Design and first results}.
\newblock \emph{\bibinfo{journal}{\pasa}} \textbf{\bibinfo{volume}{37}}, \bibinfo{pages}{e048} (\bibinfo{year}{2020}).
\newblock \eprint{2012.00747}.

\bibitem{2020PASP..132c5001L}
\bibinfo{author}{{Lacy}, M.} \emph{et~al.}
\newblock {The Karl G. Jansky Very Large Array Sky Survey (VLASS). Science Case and Survey Design}.
\newblock \emph{\bibinfo{journal}{\pasp}} \textbf{\bibinfo{volume}{132}}, \bibinfo{pages}{035001} (\bibinfo{year}{2020}).
\newblock \eprint{1907.01981}.

\bibitem{2019A&A...622A..17S}
\bibinfo{author}{{Sabater}, J.} \emph{et~al.}
\newblock {The LoTSS view of radio AGN in the local Universe. The most massive galaxies are always switched on}.
\newblock \emph{\bibinfo{journal}{\aap}} \textbf{\bibinfo{volume}{622}}, \bibinfo{pages}{A17} (\bibinfo{year}{2019}).
\newblock \eprint{1811.05528}.

\bibitem{Nayakshin2004}
\bibinfo{author}{{Nayakshin}, S.}, \bibinfo{author}{{Cuadra}, J.} \& \bibinfo{author}{{Sunyaev}, R.}
\newblock {X-ray flares from Sgr A$^{*}$: Star-disk interactions?}
\newblock \emph{\bibinfo{journal}{\aap}} \textbf{\bibinfo{volume}{413}}, \bibinfo{pages}{173--188} (\bibinfo{year}{2004}).
\newblock \eprint{astro-ph/0304126}.

\bibitem{Peters1964}
\bibinfo{author}{{Peters}, P.~C.}
\newblock {Gravitational Radiation and the Motion of Two Point Masses}.
\newblock \emph{\bibinfo{journal}{Physical Review}} \textbf{\bibinfo{volume}{136}}, \bibinfo{pages}{1224--1232} (\bibinfo{year}{1964}).

\bibitem{Eracleous1996}
\bibinfo{author}{{Eracleous}, M.}, \bibinfo{author}{{Livio}, M.}, \bibinfo{author}{{Halpern}, J.~P.} \& \bibinfo{author}{{Storchi-Bergmann}, T.}
\newblock {Elliptical Accretion Disks in Active Galactic Nuclei}.
\newblock \emph{\bibinfo{journal}{\apj}} \textbf{\bibinfo{volume}{438}}, \bibinfo{pages}{610} (\bibinfo{year}{1995}).

\bibitem{Bonnell2008}
\bibinfo{author}{{Bonnell}, I.~A.} \& \bibinfo{author}{{Rice}, W.~K.~M.}
\newblock {Star Formation Around Supermassive Black Holes}.
\newblock \emph{\bibinfo{journal}{Science}} \textbf{\bibinfo{volume}{321}}, \bibinfo{pages}{1060} (\bibinfo{year}{2008}).
\newblock \eprint{0810.2723}.

\bibitem{Goicovic2016}
\bibinfo{author}{{Goicovic}, F.~G.} \emph{et~al.}
\newblock {Infalling clouds on to supermassive black hole binaries - I. Formation of discs, accretion and gas dynamics}.
\newblock \emph{\bibinfo{journal}{\mnras}} \textbf{\bibinfo{volume}{455}}, \bibinfo{pages}{1989--2003} (\bibinfo{year}{2016}).
\newblock \eprint{1507.05596}.

\bibitem{Price2024}
\bibinfo{author}{{Price}, D.~J.} \emph{et~al.}
\newblock {Eddington Envelopes: The Fate of Stars on Parabolic Orbits Tidally Disrupted by Supermassive Black Holes}.
\newblock \emph{\bibinfo{journal}{\apjl}} \textbf{\bibinfo{volume}{971}}, \bibinfo{pages}{L46} (\bibinfo{year}{2024}).
\newblock \eprint{2404.09381}.

\bibitem{Hayasaki2016}
\bibinfo{author}{{Hayasaki}, K.}, \bibinfo{author}{{Stone}, N.} \& \bibinfo{author}{{Loeb}, A.}
\newblock {Circularization of tidally disrupted stars around spinning supermassive black holes}.
\newblock \emph{\bibinfo{journal}{\mnras}} \textbf{\bibinfo{volume}{461}}, \bibinfo{pages}{3760--3780} (\bibinfo{year}{2016}).
\newblock \eprint{1501.05207}.

\bibitem{shakura1973}
\bibinfo{author}{{Shakura}, N.~I.} \& \bibinfo{author}{{Sunyaev}, R.~A.}
\newblock {Black holes in binary systems. Observational appearance.}
\newblock \emph{\bibinfo{journal}{\aap}} \textbf{\bibinfo{volume}{24}}, \bibinfo{pages}{337--355} (\bibinfo{year}{1973}).

\end{thebibliography}

\begin{addendum}
 \item[Data and Code Availability Statement]  
The datasets generated during and/or analysed during the current study are available from the corresponding author on reasonable request. With the exception of eROSITA proprietary data, the raw data of the X-ray satellites are available from public archives.
In this work we used the following X-ray observations: \textit{Swift} ObsIDs 00014350001-00014350039, \textit{Chandra} obsID 29355, \textit{NICER} ObsIDs 7204490101-7204490160, and \textit{XMM-Newton} ObsIDs 0935191401-0935191601.
\end{addendum}

\begin{addendum}
 \item 

We thank the referees for their valuable comments, which contributed to enhancing the quality of the manuscript.
We acknowledge funding from ANID programs:  Millennium Science Initiative Program NCN$2023\_002$ (LHG, JCu, PA, SB, MLMA), Millennium Science Initiative, AIM23-0001 (LHG, FEB, MLMA), ANID-Chile BASAL CATA FB210003 (RJA, FEB, CR), FONDECYT Iniciación 11241477 (LHG), FONDECYT Regular 1211429 (JCu), FONDECYT Regular 1241422 (PA, BS), FONDECYT Regular 1231718 (RJA), FONDECYT Regular 1241005 (FEB), Fondecyt Regular grant 1230345 (CR), and Programa de Becas/Doctorado Nacional 21212344 (SB). 
CR acknowledges support from the China-Chile joint research fund.
AB acknowledged support from the Deutsche Forschungsgemeinschaft (DFG, German Research Foundation) under Germany´s Excellence Strategy – EXC 2094 – 390783311
KESF \& BM are supported by NSF AST-1831415, NSF AST-2206096 and Simons Foundation Grant 533845.
RA was supported by NASA through the NASA Hubble Fellowship grant \#HST-HF2-51499.001-A awarded by the Space Telescope Science Institute, which is operated by the Association of Universities for Research in Astronomy, Incorporated, under NASA contract NAS5-26555.
AG is grateful for support from the Forrest Research Foundation.
DH is supported by DLR grant FZK 50 OR 2406.
MK acknowledges support from DLR grant FKZ 50 OR 2307.

 We acknowledge the use of public data from the \textit{Swift} data archive through ToO proposals.  This work made use of data supplied by the UK Swift Science Data Centre at the University of Leicester.
This research has made use of data obtained from the Chandra Data Archive provided by the \textit{Chandra} X-ray Center (CXC).
 This work is supported by NASA through the \textit{NICER} mission and the Astrophysics Explorers Program and uses data and software provided by the High Energy Astrophysics Science Archive Research Center (HEASARC), which is a service of the Astrophysics Science Division at NASA/GSFC and High Energy Astrophysics Division of the Smithsonian Astrophysical Observatory.
The scientific results reported in this work are based on observations obtained with \textit{XMM-Newton}, an ESA science mission with instruments and contributions directly funded by ESA Member States and NASA.
This work is based on data from \textit{eROSITA}, the soft X-ray instrument aboard SRG, a joint Russian-German science mission supported by the Russian Space Agency (Roskosmos), in the interests of the Russian Academy of Sciences represented by its Space Research Institute (IKI), and the Deutsches Zentrum für Luft- und Raumfahrt (DLR). The SRG spacecraft was built by Lavochkin Association (NPOL) and its subcontractors, and is operated by NPOL with support from the Max Planck Institute for Extraterrestrial Physics (MPE). The development and construction of the eROSITA X-ray instrument was led by MPE, with contributions from the Dr. Karl Remeis Observatory Bamberg \& ECAP (FAU Erlangen-Nuernberg), the University of Hamburg Observatory, the Leibniz Institute for Astrophysics Potsdam (AIP), and the Institute for Astronomy and Astrophysics of the University of Tübingen, with the support of DLR and the Max Planck Society. The Argelander Institute for Astronomy of the University of Bonn and the Ludwig Maximilians Universität Munich also participated in the science preparation for eROSITA.
Part of this work is based on observations obtained with the Samuel Oschin Telescope 48-inch and the 60-inch Telescope at the Palomar
Observatory as part of the Zwicky Transient Facility project. ZTF is supported by the National Science Foundation under Grant
No. AST-2034437 and a collaboration including Caltech, IPAC, the Weizmann Institute for Science, the Oskar Klein Center at
Stockholm University, the University of Maryland, Deutsches Elektronen-Synchrotron and Humboldt University, the TANGO
Consortium of Taiwan, the University of Wisconsin at Milwaukee, Trinity College Dublin, Lawrence Livermore National
Laboratories, and IN2P3, France. Operations are conducted by COO, IPAC, and UW.
The Australia Telescope Compact Array is part of the Australia Telescope National Facility which is funded by the Australian Government for operation as a National Facility managed by CSIRO.
We thank the staff of the GMRT that made these observations possible. GMRT is run by the National Centre for Radio Astrophysics of the Tata Institute of Fundamental Research.
This scientific work uses data obtained from Inyarrimanha Ilgari Bundara / the Murchison Radio-astronomy Observatory. We acknowledge the Wajarri Yamaji People as the Traditional Owners and native title holders of the Observatory site. CSIRO’s ASKAP radio telescope is part of the Australia Telescope National Facility (https://ror.org/05qajvd42). Operation of ASKAP is funded by the Australian Government with support from the National Collaborative Research Infrastructure Strategy. ASKAP uses the resources of the Pawsey Supercomputing Research Centre. Establishment of ASKAP, Inyarrimanha Ilgari Bundara, the CSIRO Murchison Radio-astronomy Observatory and the Pawsey Supercomputing Research Centre are initiatives of the Australian Government, with support from the Government of Western Australia and the Science and Industry Endowment Fund. This paper includes archived data obtained through the CSIRO ASKAP Science Data Archive, CASDA (https://data.csiro.au).
The National Radio Astronomy Observatory is a facility of the National Science Foundation operated under cooperative agreement by Associated Universities, Inc. CIRADA is funded by a grant from the Canada Foundation for Innovation 2017 Innovation Fund (Project 35999), as well as by the Provinces of Ontario, British Columbia, Alberta, Manitoba and Quebec. 
Part of this work was carried out using the Web Interface from the ALeRCE broker.

 \item[Author Contributions]
LHG was the PI of all \textit{Swift}, \textit{Chandra}, \textit{NICER}, and \textit{XMM-Newton} ToO and DDT proposals, led the organization and writing of the paper and performed X-ray spectral analysis of \textit{Swift}, \textit{Chandra}, and \textit{XMM-Newton} data as well as the reduction and analysis of UV data. JCh performed the reduction and analysis of \textit{NICER} and \textit{XMM-Newton} data, the analysis of the evolution of the QPEs, and helped with the writing of the article. PSS discovered the optical transient, followed and improved the project and the strategy of follow-up proposals, analysed UV/optical data, obtained the probability of detecting QPEs before this discovery, and helped with the writing of the article. CR helped with the strategy of follow-up proposals and actively contributed to all intermediate results before submission. JCu, BM and KESF developed theoretical models and helped with the writing of the article. 
PA calculated the integrated flux from accretion disk models to compare with the data.
AR, RA, PA, EK, ZL and AM contributed to interpretation of results and discussed optimal data reduction procedures. PA and BS provided the ZTF light curves. GB helped with archival radio data, and AG was the PI of the ATCA DDT request and analysed these data. ZA and KG scheduled and oversaw the \textit{NICER} observations. 
PB contributed with the time resolved \textit{eROSITA} data.
MB reduced the \textit{Chandra} data and helped with the analysis. 
RJA, AB, FEB, SB, GCR, MK, PL, MLMA, and MS provided input to the manuscript text and interpretation. 
CR, FEB, PA, AR, RA, EK, ZL, and AM provided feedback during long-term follow-up campaigns. 
All authors contributed to discussing data analysis
and results, read the article since first draft, and helped improving it during the whole process.

 \item[Author Information] The authors declare that they have no
competing financial interests. Correspondence and requests for materials
should be addressed to L.H.G. \\
(email: lorena.hernandez@uv.cl).
\end{addendum}

\newpage
\pagebreak

\captionsetup[figure]{name=Extended Data Figure}
\captionsetup[table]{name=Extended Data Table}

\setcounter{section}{0}
\setcounter{figure}{0}
\setcounter{table}{0}

\noindent \textbf{\huge Methods\label{methods}}

\section{The optical/UV transient ZTF\,19acnskyy}

For two decades, the galaxy SDSS1335+0728 showed no optical or infrared variations; however, in December 2019 (MJD 58830), the ZTF alert stream detected significant nuclear variability (as ZTF19acnskyy and later as ZTF22abyhaut). 
From its original name in ZTF, we call this event ``Ansky'', for easier identification of the source.
On this date, Ansky increased its flux by $\sim20\%$ (with respect to the ZTF $g$ band template). It reached its peak in May 2020 (MJD 58991), and it has been exhibiting a slow decay since then, showing stochastic variations for more than 1,680 days. In 2020, the ALeRCE (Automatic Learning for the Rapid Classification of Events; \citeExtra{forster2021}) broker light curve classifier \citeExtra{Sanchez-Saez21a} classified the source as an AGN, which motivated a spectroscopic and photometric follow-up campaign with different facilities to understand its nature, which was presented in \cite{Sanchez24}. 

The source was observed with \textit{Swift}/UVOT (July 2021, July 2022, July 2023, February/March/April 2024). The analysis of the data revealed that in 2021, the UV flux was four times larger than the values reported by  GALEX in 2004. Moreover, the analysis of the ZTF light curves shown that although the optical flux has declined in the last four years, the slope of this decay is far from what is expected for classical TDEs, with power-low indices close of -0.17 and -0.14 in the $g$ and $r$ bands, respectively.   

A spectroscopic follow-up campaign was conducted using SOAR/Goodman (July 2021 and January 2024; 1$\arcsec$ slit), VLT/XSHOOTER (July 2022, $0.5\arcsec$ slit), and Keck/LRIS (July 2023; $0.7 \arcsec$ and $1.5\arcsec$ slits). Moreover, archival data prior to the first ZTF alert, SDSS1335+0728 was obtained from SDSS (2007; $3\arcsec$ fibre) and LAMOST (2015; $3.5\arcsec$ fibre). None of the archival and new spectra showed broad emission lines (BEL), or Bowen fluorescence (BF) emission lines (recently found in flaring AGNs; \citeExtra{Trakhtenbrot19NatAs}), but did exhibit narrow emission line (NEL) and continuum variability. \cite{Sanchez24} presented the BPT diagram (Baldwin, Phillips, and Terlevich; \citeExtra{Baldwin81}) evolution of Ansky, showing that the LAMOST spectrum is classified as star-forming, while the SDSS spectrum as composite. The first two follow-up observations (Goodman 2021 and XSHOOTER 2022) fall near the limit between composite-AGN and Star-Forming-AGN, indicating a more energetic ionisation state. The two most recent observations (LRIS 2023 and Goodman 2024) show a continuing evolution towards the AGN regions of the BPT diagram. The most drastic variations appear from the [OIII] NEL, which suggests that the Narrow Line Region (NLR) has had enough time to react to the increased ionising continuum after more than 3.6\,years of activity (time between the first alert and the LRIS observation). 

In addition, from the latest WISE data release, \cite{Sanchez24} found that the mid-infrared flux has risen more than two times between June 2022 and June 2023, and that the W1$-$W2 WISE colour has become redder. The recent MIR evolution cannot be explained by a pre-existing AGN-like dusty torus (a shorter timescale for the MIR flux rise would be expected; \citeExtra{Lyu19}), which suggests that either the dust reprocessing is happening in a much larger dusty structure (with sizes of few light years), or that an AGN-like torus is under formation.

From the follow-up campaign and the ZTF and WISE observations, \cite{Sanchez24} proposed two hypotheses for the origin of these variations: 1) a `turning on' AGN \citeExtra{Arevalo24}, or 2) an exotic TDE. If the former, it represents one of the strongest cases of an AGN observed while ``activating''. If the latter, it would correspond to the longest and faintest TDE ever observed or to another class of still unknown exotic nuclear transient.

\begin{figure*}[t!]
 \centering
 \hspace{-2cm*}\includegraphics[width=1.1\textwidth]{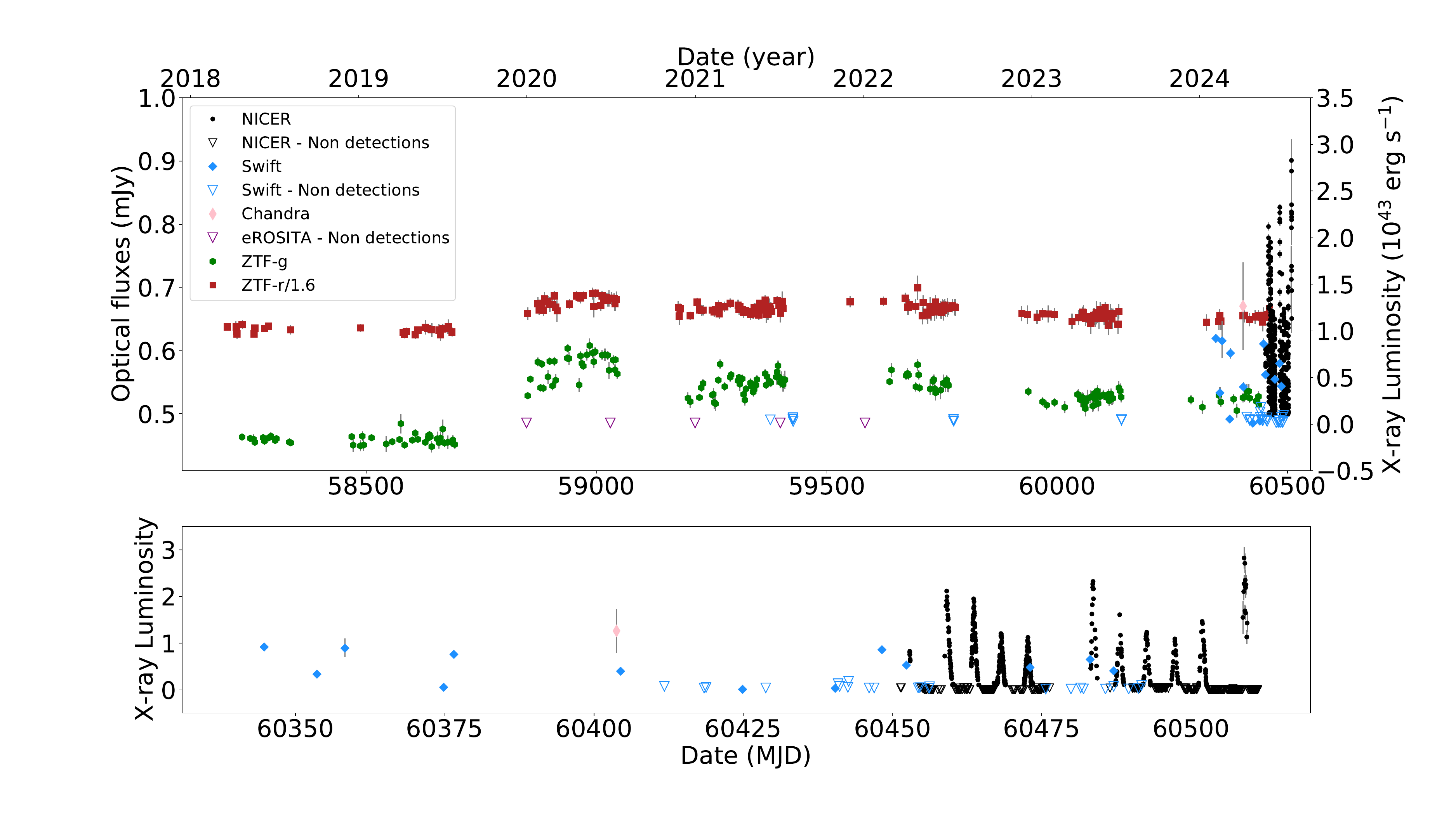}
 \caption{Long term light curve of ZTF\,19acnskyy (Top): between 2018-2024. It includes optical data from ZTF in the $g$ (green circles) and $r$ (red squares, divided by a factor 1.6) bands, with the total fluxes in units of mJy (shown in the left axis), and X-ray data from \textit{NICER} (detections as black circles, non detections as black triangles), \textit{Swift}/XRT (detections as blue diamonds, non detections as blue triangles), \textit{Chandra} (pink thin-diamond), and \textit{eROSITA} non detections (as purple triangles), with the X-ray luminosities in units of erg s$^{-1}$ (shown in the right axis); (Bottom): zoom-in to X-ray data on dates after February 2024, when X-ray emission was detected for the first time with \textit{Swift}/XRT. The X-ray luminosity is given in units of 10$^{43}$erg s$^{-1}$. }
\label{fig:ztf}
\end{figure*}

\section{Observations and data analysis}

\subsection{X-ray data}\label{sec:sample_xray}
We obtained data from several X-ray telescopes in our campaign to monitor the QPEs in Ansky. Here we describe the data reduction and analysis procedures. All the spectral fits were performed using \texttt{XSPEC} v12.13.1 and the galactic absorption was fixed to $N_H=2.6\times 10^{20}$ cm$^{-2}$. No intrinsic absorption was required in any of the fits.

\subsubsection{Swift}\label{sec:swift}  Observations with the \swift\ X-ray Telescope (XRT, \citeExtra{Burrows05}), onboard the Neil Gehrels Swift Observatory, were taken in the Photon Counting (PC) mode through ToO proposals (IDs. 15797, 15870, 17585, 19071, 19784, 20041, 20083, 20177, 20363, 20405, 20489, and 20623). The observations were carried out between June 2021 and July 2024, and the exposure times of all the observations sum up to $\sim$50 ksec.
The first X-ray detection was on February 4, 2024 through ToO ID 20041 \citep{Hernandez24}.

We performed the data reduction following standard routines described by the UK Swift Science Data Centre (UKSSDC). Calibrated event files were produced using the routine {\sc xrtpipeline}, accounting for bad pixels and effects of vignetting, and exposure maps were also created. Source and background spectra were extracted from circular regions with 30 arcsec and 50 arcsec radius, respectively. The background was a region free of sources located close to the nucleus. The {\sc xrtmkarf} task was used to create the corresponding ancillary response files. The response matrix files were obtained from the HEASARC CALibration DataBase. The spectra were grouped to have a minimum of 5 counts per bin using the {\sc grppha} task. 

The spectra were fit with the \texttt{XSPEC} model \texttt{tbabs$\times$zbbody}, yielding temperatures in the range kT=[30-100] eV. The energy range 0.3-1 keV was used for the spectral using with C-stat statistics \citeExtra{cash1979}.

As a double check, count rates and upper limits were retrieved from the living Swift-XRT point source catalogue  \citeExtra{LSXPSCatalogue2023}, and converted into flux using WebPIMMS and a blackbody model using a temperature of 100 eV. The flux upper limits are in the range between 1.2$\times$10$^{-13}$ and 1.4$\times$10$^{-12}$ erg s$^{-1}$cm$^{-2}$.

Fig.~\ref{fig:lc} and Extended Data Fig.~\ref{fig:ztf} include flux measurements of the \textit{Swift} data.
The fluxes are in the range Flux(0.3-2 keV) = [0.05-7.5]$\times$10$^{-12}$erg s$^{-1}$cm$^{-2}$, which corresponds to observed luminosities in the range of L(0.3-2 keV) = [0.1-9.2]$\times$10$^{42}$erg s$^{-1}$.

\subsubsection{Chandra}\label{sec:chandra}

A \textit{Chandra} \citeExtra{Weisskopf2000} Director’s Discretionary Time (DDT) observation was obtained on April 3, 2024 (MJD 60403), with an exposure of 2 ks using the ACIS-S instrument and the main aim to obtain an accurate position of the source.  We did source detection by running {\tt wavdetect} with wavelet scales of 1, 2 and 4$\arcsec$ on the 0.5--8 keV image. This yielded 19 X-ray sources including ZTF\,19acnskyy which was well detected with a count rate of 0.012 ct\,s$^{-1}$ in the 0.5--2 keV band. We then cross-matched these sources with the Gaia DR3 catalog with a search radius of 1$\arcsec$, finding two matches including ZTF\,19acnskyy and the nucleus of SDSS1335+0728. The mean offset between the X-ray sources and the Gaia ones was 0.26" which confirms that ZTF\,19acnskyy is coincident with the nucleus of SDSS1335+0728.
 
 Source and background spectra were extracted from 2$\arcsec$ and 10$\arcsec$ radii, respectively using the {\sc specextract} task to extract the spectral products. We grouped the spectrum with a minimum of 1 count per bin with {\tt grppha} and fitted it in \texttt{XSPEC} using the C-statistic and the background subtracted. We fitted the model \texttt{tbabs$\times$zbbody} with the absorption fixed to $N_{\rm H}=2.6\times10^{20}$ cm$^{-2}$ as done for the \textit{XMM-Newton} data and the resulting spectral parameters were $kT=88^{+17}_{-14}$ eV. The observed 0.3--2 keV flux was 6.7$^{+10}_{-3.9}\times10^{-12}$ which implies an X-ray luminosity L(0.3-2 keV) = 8.7$^{+13}_{-5.1}\times$10$^{42}$ erg s$^{-1}$ when assuming a redshift of 0.024.

\subsubsection{NICER}\label{sec:nicer} 
The \textit{NICER} X-ray Timing Instrument \citeExtra{Gendreau16} observed Ansky for a total of 278~ks across 60 Target of Opportunity (ToO) observations (ObsIDs 7204490101-7204490160) from May 19-July 20, 2024. The data were processed using \texttt{HEAsoft} v6.33 and \texttt{NICERDAS} v12. We followed the time-resolved spectroscopy approach to reliably estimate the background-subtracted light curve outlined in \cite{Chakraborty24}, which we briefly summarize here. First we use \texttt{nimaketime}, with unrestricted undershoot (\texttt{underonly\_range}=*-*) and overshoot rates (\texttt{overonly\_range}=*-*), as well as per-FPM and per-MPU autoscreening disabled. We then split the intervals produced by \texttt{nimaketime} into Good Time Intervals (GTIs) of 200 seconds to allow estimation of the time-varying background. In each GTI, we manually discard focal plane modules (FPMs) with 0-0.2 or 5-15 keV count rates $>4\sigma$ higher than the average across all GTIs within the ObsID to discard intervals of significant overshoot/undershoot rate. After screening the event lists, we used the \texttt{SCORPEON}\footnote{\href{https://heasarc.gsfc.nasa.gov/lheasoft/ftools/headas/niscorpeon.html}{\texttt{https://heasarc.gsfc.nasa.gov/lheasoft/ftools/\\headas/niscorpeon.html}}} template-based background model to estimate the contribution from astrophysical and non X-ray backgrounds separately for each GTI. We fit the entire broadband (0.2-15 keV) array counts with the PyXspec\footnote{\href{https://heasarc.gsfc.nasa.gov/docs/xanadu/xspec/python/html/index.html}{\texttt{https://heasarc.gsfc.nasa.gov/docs/xanadu/xspec/\\python/html/index.html}}} interface to \texttt{XSPEC} \citeExtra{Arnaud96}. We leave the Solar Wind Charge Exchange (SWCX) oxygen emission line normalizations free to vary to account for partially ionized oxygen fluorescence from the solar wind (which is particularly important during \nicer\ dayside GTIs). Along with the \texttt{SCORPEON} background, we fit each GTI with a source model represented by \texttt{tbabs}$\times$\texttt{zbbody} typical of QPEs. We estimate the model luminosity of the source model only to create the light curve presented in Fig.~\ref{fig:lc}. We consider a source ``detection" as any GTI in which the blackbody normalization is $>$1$\sigma$ inconsistent with zero, i.e. a non-background component is required by the fit at the $1\sigma$+ level. The flux upper limits are in the range 2.6$\times$10$^{-18}$ and 3.5$\times$10$^{-13}$ erg s$^{-1}$cm$^{-2}$.

We also perform spectroscopy with coarser time-resolution to track the evolution of the QPE bolometric luminosity ($L_{bol}$) and temperature  ($kT$), from which we inferred the blackbody radius ($R_{bb}$) over the different phases of each flare (Fig.~\ref{fig:hyst}). We followed the same filtering and background estimation procedures described above, then combined all GTIs within 8 bins per QPE based on the relative intensity compared to the peak. We chose the bin cutoffs during the QPE rise to correspond to 1-25\% $F_{peak}$, 25-50\%, 50-75\%, 75-100\%, and the reverse for the decline. Following this, we used the routine \texttt{niobsmerge} to merge all ObsIDs falling within these bounds, then fit with the same time-resolved spectroscopy procedure described above. We fit over a large energy range (0.25-10 keV) for data taken in orbit night, and a restricted range (0.38-10 keV) during orbit day, to allow accurate estimation of the background. We used the default binning chose of \texttt{grouptype=optmin} with \texttt{groupscale=10} \citeExtra{Kaastra16}, and fit with the Cash statistic (cstat). After fitting each spectrum separately, we took the average of each bin across all QPEs to compute the average evolution shown in Fig.~\ref{fig:hyst}. The gaussian absorption component is also observed in \textit{NICER} data, confirming its astrophysical (rather than instrumental) origin.

\subsubsection{XMM-Newton}\label{sec:xmm}
\begin{figure*}[t!]
 \centering
 \includegraphics[width=0.33\textwidth]{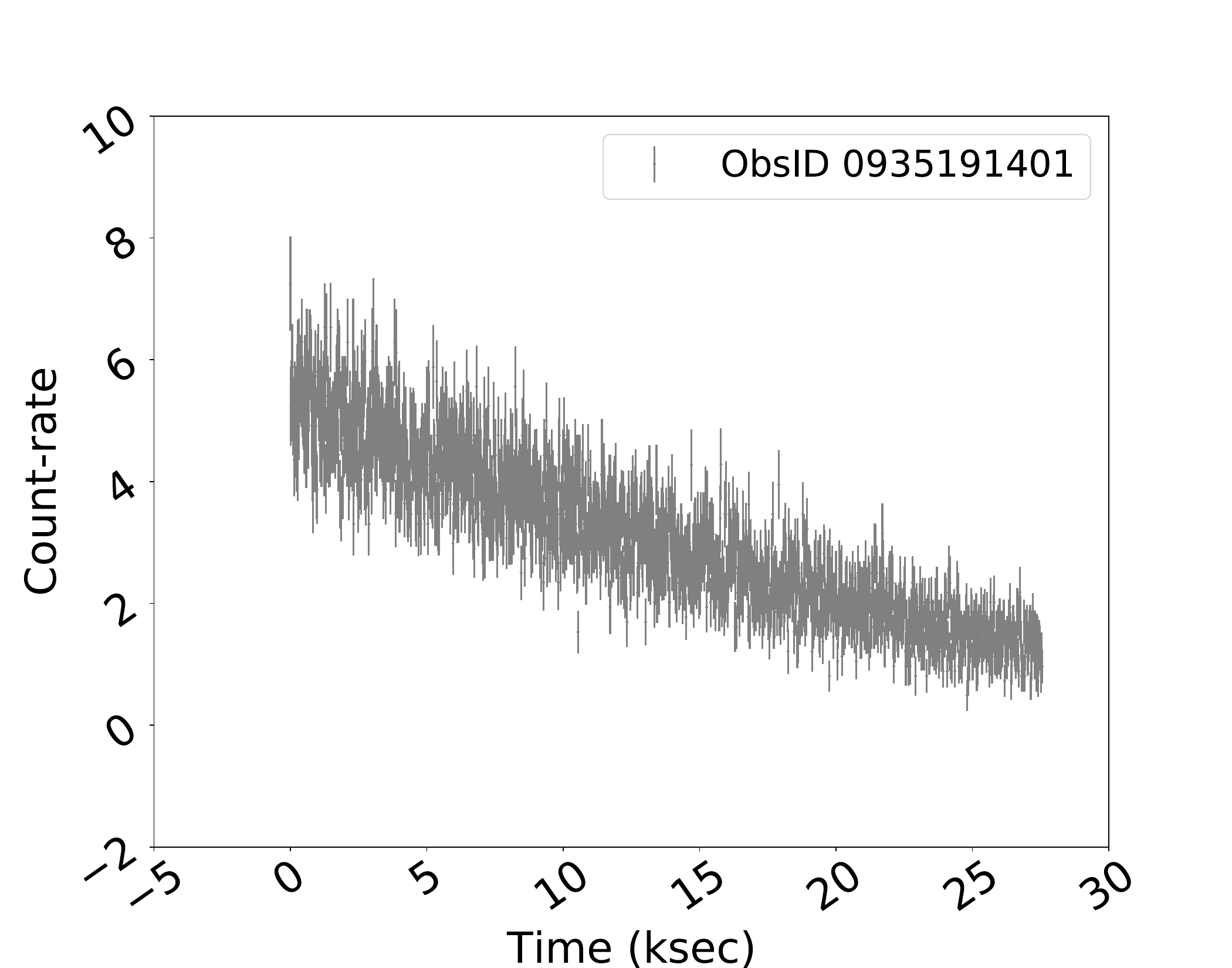}
 \includegraphics[width=0.33\textwidth]{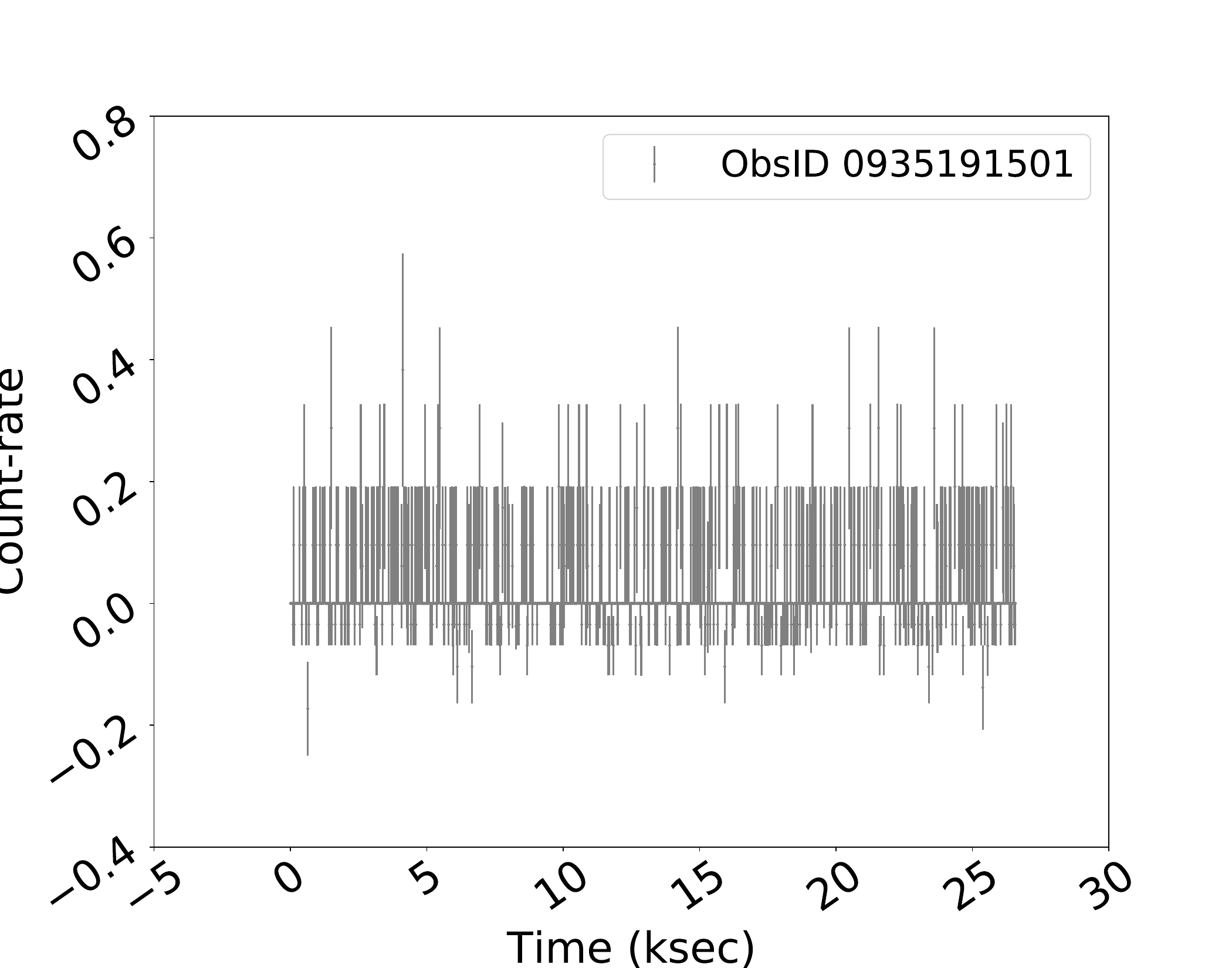}
 \includegraphics[width=0.33\textwidth]{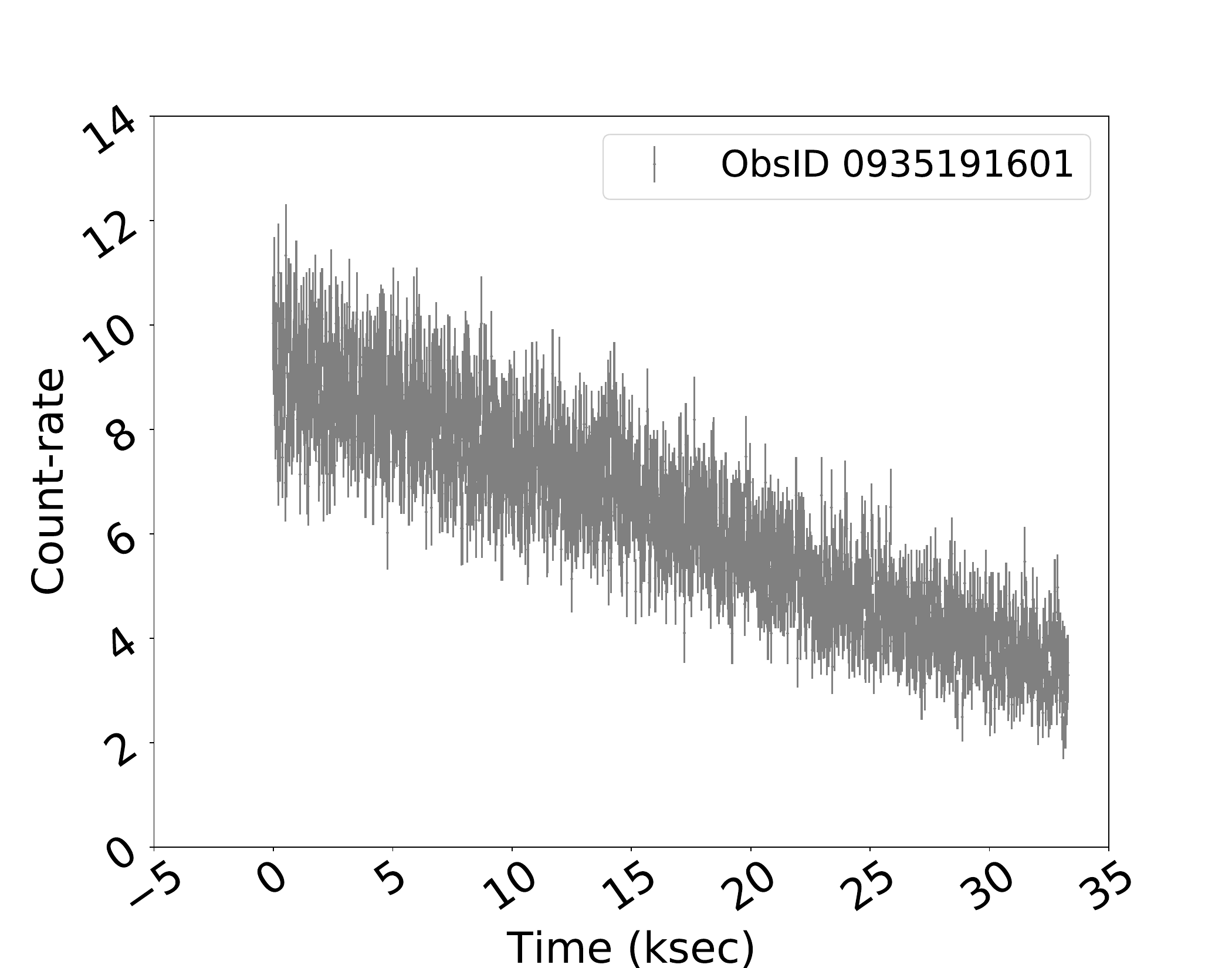}
 \caption{\textit{XMM-Newton} light curves of ObsID 0935191401 (left), 0935191501 (middle), and 0935191601 (right). The light curves are background-subtracted and grouped into 20-second bins.
 }
\label{fig:xmm_lc}
\end{figure*}

\textit{XMM-Newton} observed Ansky for a total of 87.4~ks across three Director's Discretionary Time (DDT) observations (ObsIDs 0935191401-0935191601) on June 27, June 29, and July 1, 2024. The data were reduced using XMM SASv21.0.0 and HEASoft v6.33. We restrict our analysis to the EPIC-pn instrument due to its higher count rate compared to the MOS CCDs. Source products were extracted within a circle of 33" radius, while background was extracted from an annular region free of sources centered on the source with an inner radius 40" and other radius 60". For ObsIDs 0935191401 \& 0935191601 the detector experienced significant pile-up toward the core of the point-spread function, so we excised the inner 5" to mitigate its effects in our spectral and temporal analysis. Source and background rates were then extracted, and grouped into 20-second bins to produce the background-subtracted light curves in Extended Data Fig.~\ref{fig:xmm_lc}. We extracted spectra with \texttt{evselect} with a minimum of 20 counts per bin for ObsIDs 0935191401 \& 0935191601 (eruptions) and fit with $\chi^2$ statistics, and 1 count per bin for ObsID 0935191501 (quiescence) for which we fit with the C-statistic. For ObsIDs 0935191401 and 0935191601, we also extracted phase-resolved spectra with \texttt{evselect} by dividing the total exposure into three equal-duration segments. We analyzed the background subtracted spectra using data from 0.3 to $\sim$1.1 keV (as the source is background-dominated at higher energies). 

\begin{figure}[t!]
 \centering
 \includegraphics[width=0.48\textwidth]{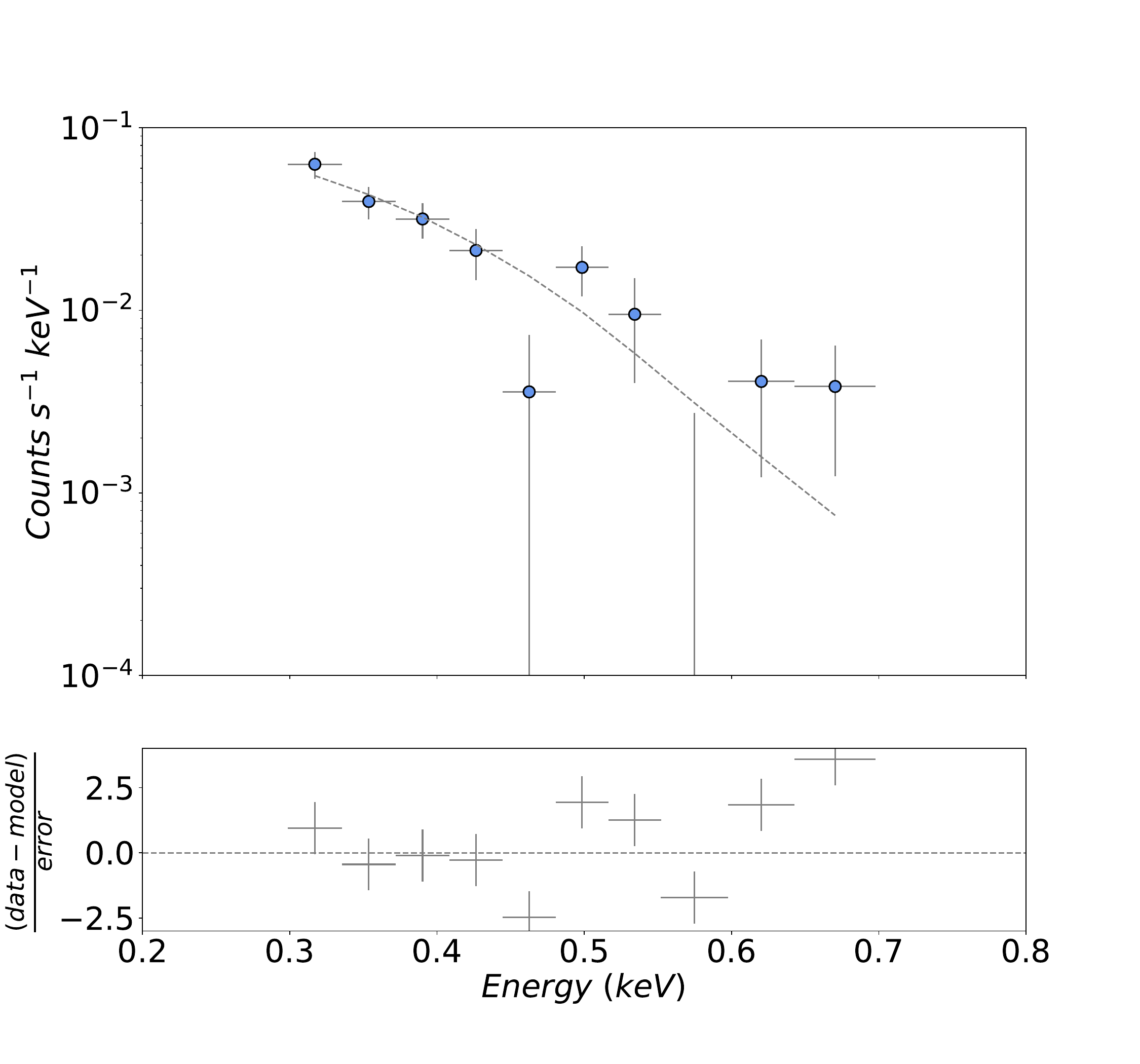}
 \caption{Spectral fit of ObsID 935191501 of \textit{XMM-Newton}. The data were obtained during the quiescence state. The model is {\sc tbabs$\times$zashift$\times$diskbb}. Strong residuals are observed around 0.7-1.0 keV band in bins 1 and 2.
 }
\label{fig:xmm_quiescence}
\end{figure}

The \textit{XMM-Newton} light curves are presented in Extended Data Fig.~\ref{fig:xmm_lc}. As ObsID 0935191501 was obtained during the quiescence, we use the spectral model \texttt{tbabs$\times$zashift$\times$diskbb}.
We used an additional component to check for the presence of intrinsic absorption using the model \texttt{zwabs}, and used the F-test to check for improvements in the spectral fit.
No intrinsic absorption was required by the data. This model resulted in a temperature of kT = 51[39-59] eV, and a flux of 2.0$\times$10$^{-14}$[1.9$\times$10$^{-15}$ - 2.1$\times$10$^{-14}$] erg s$^{-1}$cm$^{-2}$, or a luminosity of 2.8$\times$10$^{40}$[6.4$\times$10$^{38}$-3.0$\times$10$^{40}$] erg s$^{-1}$ in the 0.3--2 keV energy band. Extended Data Fig.~\ref{fig:xmm_quiescence} shows the spectral fit and its residuals.

We used a disk model to fit ObsID 0935191501 because the emission during quiescence is likely generated form a different physical region (the accretion disk, \citep[e.g.,][]{Miniutti19, Giustini20}) compared to the QPEs (shocked ejecta, \citep[e.g.,][]{Linial23, Franchini23}). 
We recall that there is no emission above $\sim$1 keV, thus the emission is consistent with thermal emission instead of the ubiquitous corona observed in AGN that emit above this energy range and is fitted by a power-law model. 
However, little is known about the formation of the X-ray corona, so it could be that the same conditions that lead to not having a broad line region also results in no corona. 1ES 1927+654 is an example of the drastic transformation of the X-ray properties in an AGN, where the X-ray corona was destroyed within a changing-look event \citeExtra{ricci2020}.

\begin{figure}[t!]
\centering
\includegraphics[width=0.5\textwidth]{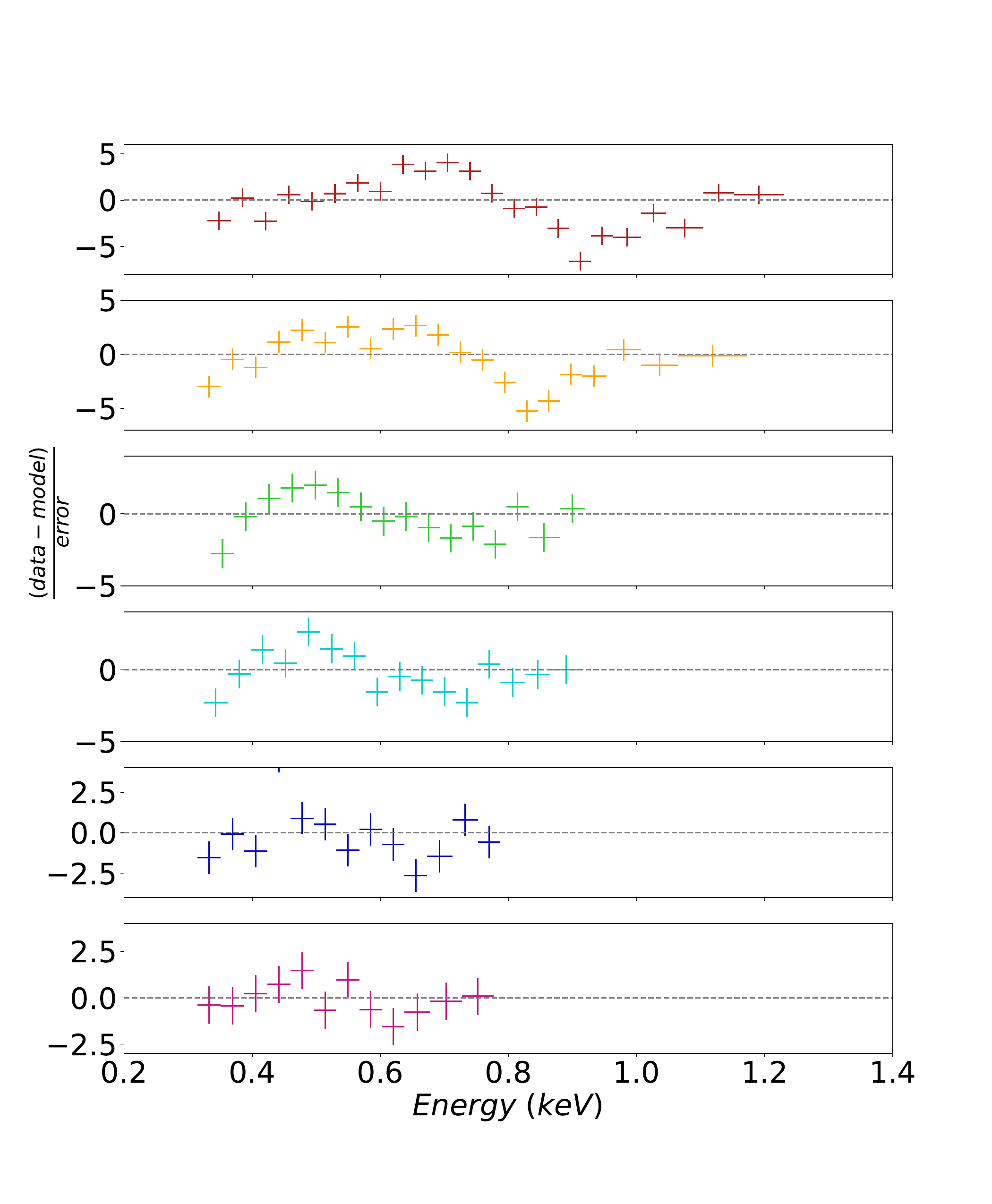}
\caption{Residuals of the spectral fits of the phase-resolved \textit{XMM-Newton} data using the model {\sc tbabs*zbbody}. From top to bottom, the observations correspond to bins 1, 2, and 3 of ObsID 035191601, and bins 4, 5, and 6 of  ObsID 035191401.}
\label{fig:fit_xmm_nogauss}
\end{figure}

\begin{figure}[t!]
\centering
\includegraphics[width=0.5\textwidth]{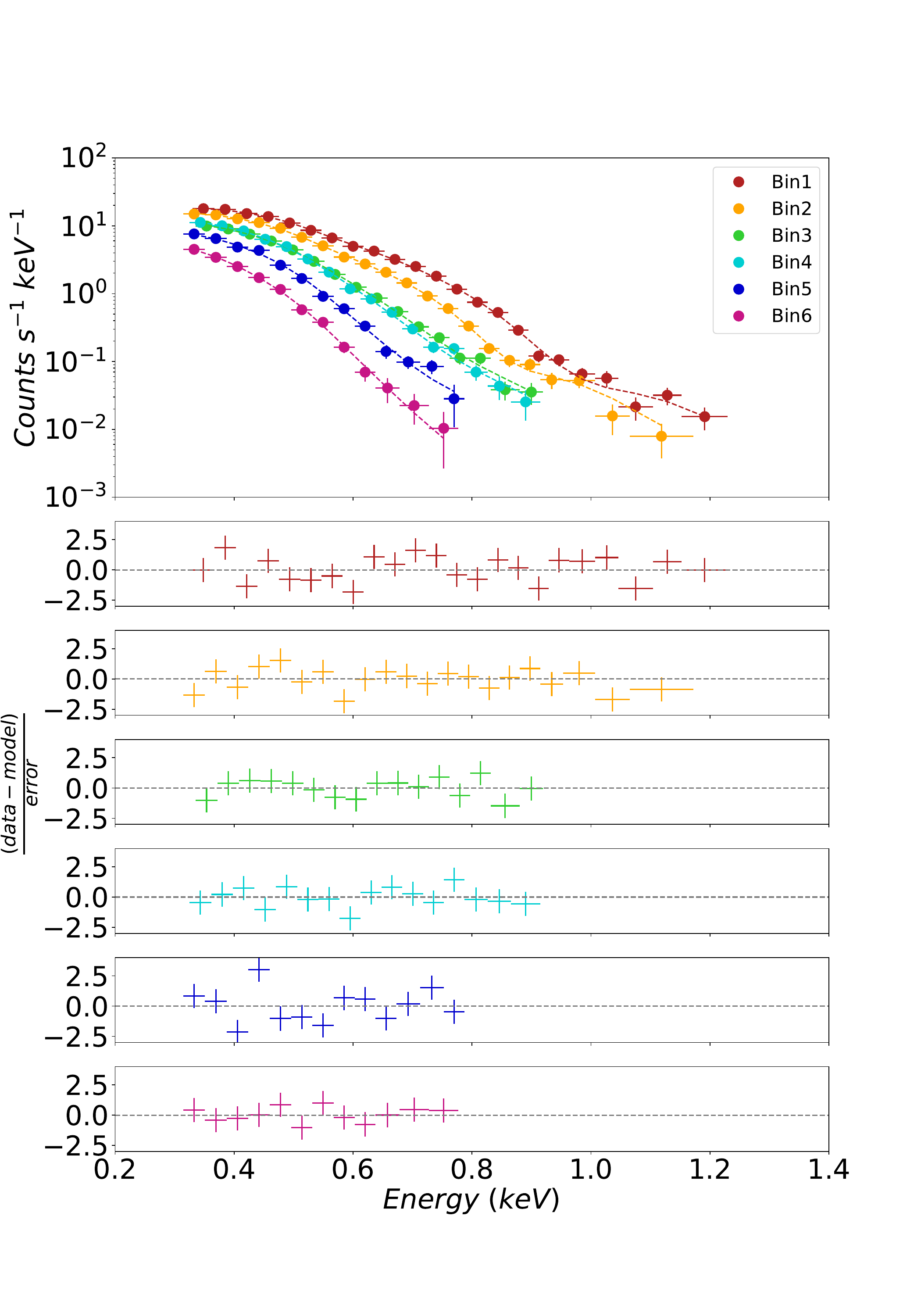}
\caption{Spectral fits (top panel) and residuals (bottom panels) of the phase-resolved \textit{XMM-Newton} data using the model {\sc tbabs*gabs*zbbody}. From top to bottom, the observations correspond to bins 1, 2, and 3 of ObsID 035191601, and bins 4, 5, and 6 of  ObsID 035191401. }
\label{fig:fit_xmm}
\end{figure}

The phase-resolved spectra correspond to bins 1, 2, and 3 of ObsID 035191601, and bins 1, 2, and 3 (4, 5, 6) of  ObsID 035191401. 
We initially used the \texttt{XSPEC} model \texttt{tbabs$\times$zbbody} to fit the QPE emission. For the higher S/N spectra of \xmm\ ObsIDs 0935191401 and 0935191601. As shown in Extended Data Fig. \ref{fig:fit_xmm_nogauss}, which shows the residuals of this model for each bin, this fit produces significant deficit in the counts/model ratio around $\sim 1$ keV, with reduced $\chi^2 > 3$.
Accordingly, we use \texttt{tbabs$\times$gabs$\times$zbbody} to model the feature with a gaussian absorption profile.
Here we report only the need of this feature and a detailed time-resolved analysis of this absorption feature and physical interpretation of its properties is presented in a follow-up paper \citeExtra{Chakraborty25}.
The same procedure was adopted to check for intrinsic absorption, but this component was not required by the data.
Results of the spectral fitting are reported in Table~\ref{tab:spectral_fit_xmm} and Extended Data Fig.~\ref{fig:fit_xmm}. From this fit, we obtained a range of temperatures of kT = [45-76] eV and Flux(0.3-2 keV) = [1.5-8.0]$\times$10$^{-12}$erg s$^{-1}$cm$^{-2}$, which correspond to luminosities in the range of L(0.3-2 keV) = [2-11]$\times$10$^{42}$erg s$^{-1}$.
We tested a model with a same Gaussian line for all the spectra at E$\sim$0.9 keV, and the spectral parameters do not change. Searching for variations in the Gaussian feature is out of the scope of this work and will be presented in a forthcoming paper.

\begin{table*}[h!]
\caption[]{Spectral parameters of \textit{XMM-Newton} spectra.  Phase resolved spectral fitting discussed in Section \ref{sec:xmm} and shown in Extended Data Fig. \ref{fig:fit_xmm}. The model \texttt{tbabs*gabs*zbbody} was used with frozen galactic $N_H=2.6\times 10^{20}$ cm$^{-2}$. The table lists the parameters for the absorption feature: line energy in keV, Line E, the line width in keV, $\sigma$, and line depth in keV, Strength; the parameters of the blackbody: temperature in eV, kT, and its normalization; the chi$^2$ and degrees of freedom, d.o.f., and the fluxes and luminosities in the 0.3-2 keV energy band. Bins 1, 2, 3 are derived from obsID 601, while bins 4, 5 and 6 are derived from obsID 401.  Errors are estimated at 90\%.}
\label{tab:spectral_fit_xmm}
\begin{center}
\begin{tabular}{lcccccc}
\hline
\hline
\noalign{\smallskip}
 & bin1 &  bin2 &	bin3	& bin4 & bin5 & bin6 \\ 
\hline

Line E &  0.96[0.94-0.99] & 0.85[0.83-0.89] & 0.76[0.73-0.80] & 0.73[0.70-0.76] & 0.7[0-0.73] & 0.7[0-0.75] \\
$\sigma$ & 0.08[0.06-0.10] & 0.05[0.03-0.09] & 0.10[0.07-0] & 0.10[0.06-0] & 0.10[0.08-0] & 0.010[0.04-0] \\
Strength & 0.24[0.18-0.32] & 0.15[0.11-0.23] & 0.16[0.10-0.22] & 0.17[0.09-0.22] & 0.28[0.25-0.34] & 0.21[0.14-0.36]\\
kT (eV) &  76.0[75.3-76.8] & 68.3[67.6-69.2] & 62.0[60.3-63.9] & 59.4[57.5-61.1] & 52.9[51.1-55.0] & 45.5[43.5-49.0] \\
Norm & 3.51[3.42-3.61]e-4 & 3.51[3.39-3.63]e-4 & 3.00[2.76-3.27]e-4 & 3.61[3.53-3.99]e-4 & 2.92[2.83-3.45]e-4 & 3.33[3.18-4.50]e-4 \\
$\chi^2$/dof & 25.0/18 & 16.0/16 &  8.7/11 & 9.0/11 & 22.5/8 & 4.2/7 \\
F(0.3-2 keV) & 7.98[7.89-8.07] &  6.27[6.19-6.35] & 4.06[3.93-4.16] & 4.27[4.09-4.37] &  2.61[2.14-2.80] & 1.49[0.91-1.58]  \\
(10$^{-12}$ erg s$^{-1}$cm$^{-2}$) & \\
L(0.3-2 keV) & 10.77[10.63-10.88] & 8.52[8.40-8.63] & 5.57[5.35-5.69] & 5.87[5.62-6.03] &  3.61[2.98-3.85] & 2.11[1.37-2.24]  \\
(10$^{42}$ erg s$^{-1}$) & \\
\noalign{\smallskip}
\hline
\noalign{\smallskip}
\end{tabular}
\end{center}
\end{table*}

%
%

\subsubsection{eROSITA}\label{sec:erosita}
eROSITA observed the field of Ansky in five eRASS (MJD 58848 with eight epochs; 59030 with six epochs, 59214 with six epochs, 59399 with seven epochs, and 59583 with six epochs). The total baseline of each series of observations ranges between 0.833 and 1.167 days. The source was not detected in any of the five eRASS epochs, nor in the stacked image that includes the first four eRASSes. Upper limits from the stacked images are provided in \citeExtra{Tubin24}. In the most sensitive eROSITA band with $0.2-2.3$~keV, the reported upper limit in the field of Ansky is $3.98\times10^{-14}$ erg s$^{-1}$ cm$^{-2}$. 

\smallskip
\subsection{Optical/UV data}
\subsubsection{Zwicky Transient Facility}
The ZTF \citeExtra{graham2019, bellm2019, masci2019, masci2023} has been surveying the Northern sky every three days in the g, r and i optical filters since 2018 
and offers different services, including 1) a public alert system, for real-time, time-domain science,
2) Data Releases (DRs) every two months, including photometry measurements on the science images 
and 
3) the Forced Photometry Service on-demand and per source, including point spread function (PSF) photometry measurements on the difference images (reference-subtracted science images).   

As the PSF Forced Photometry light curves may have underestimated photometric errors (see \citeExtra{masci2023}), we constructed our own aperture Forced Photometry light curves using the same strategy described in Arevalo et al. (in preparation). In summary, we retrieved the publicly available difference images in the field of Ansky (ZTF field 476) and kept only those with good quality (i.e. with ZTF metadata $\text{infobits}=0$, $\text{maglimit}>20$ AB mag, $\text{seeing }< $4$\arcsec$), and using at most one observation per night. The photometry was forced on the difference images at the location of all sources detected in the reference image of the field, using an aperture of 4$\arcsec$. We then added the flux measured in the reference image, also in an aperture of 4$\arcsec$, to all the detected sources in order to obtain total fluxes. We corrected the photometric errors by selecting a set of non-variable stars in the field and modeled the variance of the photometry as a function of flux in each epoch. From this, we computed the variance per epoch at the flux level of Ansky in each band, and then we added it in quadrature to the errors measured from the difference images. The final aperture forced photometry light curve obtained for Ansky in the $g$ and $r$ bands is presented in total flux
in Extended Data Figure~\ref{fig:ztf}.

\subsubsection{\textit{Swift}/UVOT}
The Ultraviolet and Optical Telescope (UVOT, \citeExtra{2005SSRv..120...95R}) onboard the Neil Gehrels \textit{Swift} Observatory has six primary photometric filters: V (centred at 5468 \AA), B (at 4392 \AA), U (at 3465 \AA), UVW1 (at 2600 \AA), UVM2 (at 2246 \AA) and UVW2 (at 1928 \AA). We obtained data in the six filters simultaneously with the X-ray observations on each epoch.

The {\sc uvotsource} task within software HEASoft version 6.33 was used to perform aperture photometry using a circular aperture of radius 5 arcsec centred on the coordinates of Ansky. The background region was selected free of sources adopting a circular region of 20 arcsec close to the nucleus. 

There are certain areas in the UVOT detector where the throughput is lower than for the rest of the detector. Ansky is located in these low sensitivity areas
in some of the observations. Moreover,
observations taken between August 2023 and the start of April 2024  were affected by spacecraft jitter \citeExtra{2023GCNswift, 2024GCNswift}. This effect causes the sources in the UVOT images to appear elongated instead of point-like sources, underestimating the measured magnitudes by 0.1-0.3 mag. The UVOT team advised that the data can be used but with caution (private communication).

\begin{figure*}[t!]
 \centering
 \begin{tabular}{cc}
\includegraphics[width=0.5\textwidth]{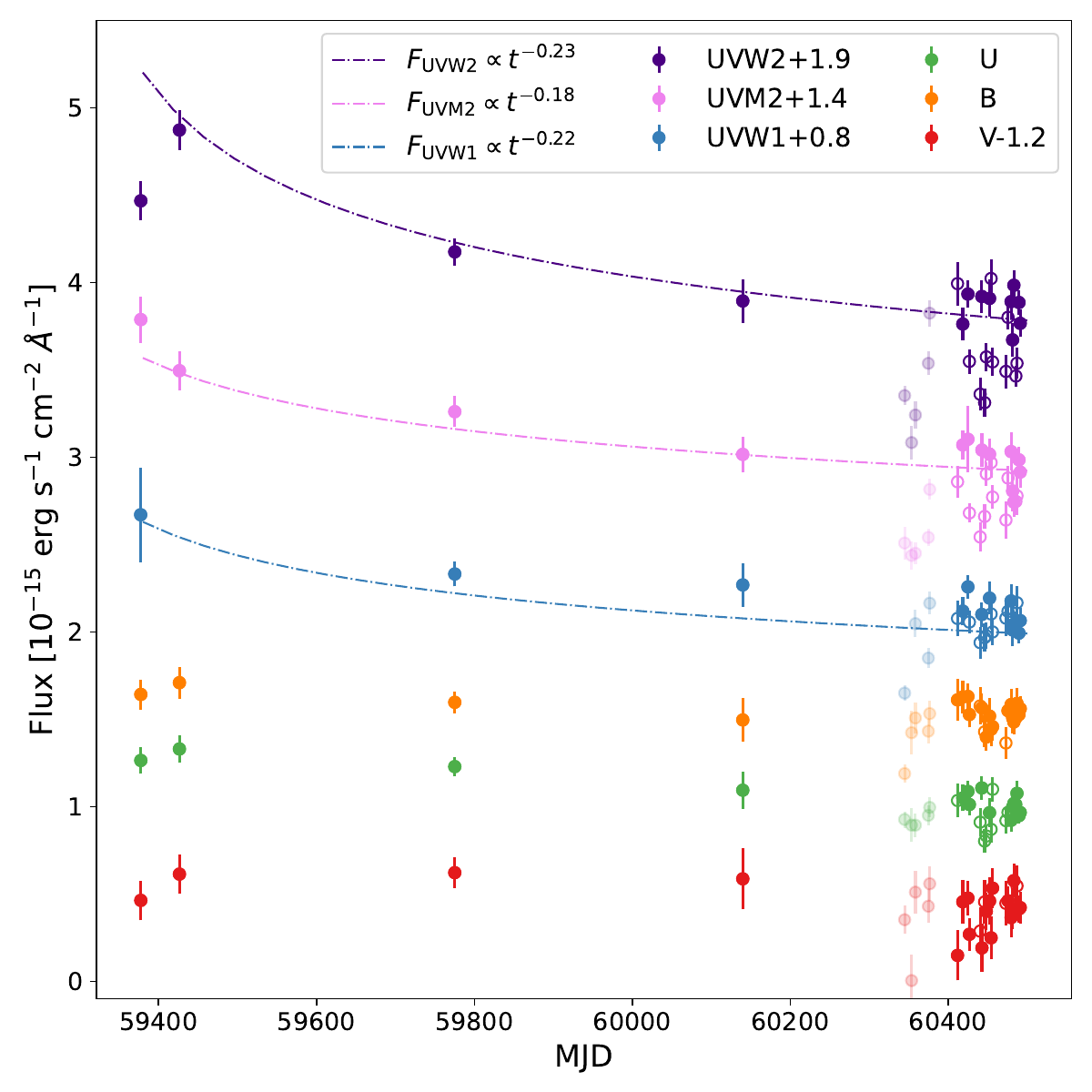}&
\includegraphics[width=0.5\textwidth]{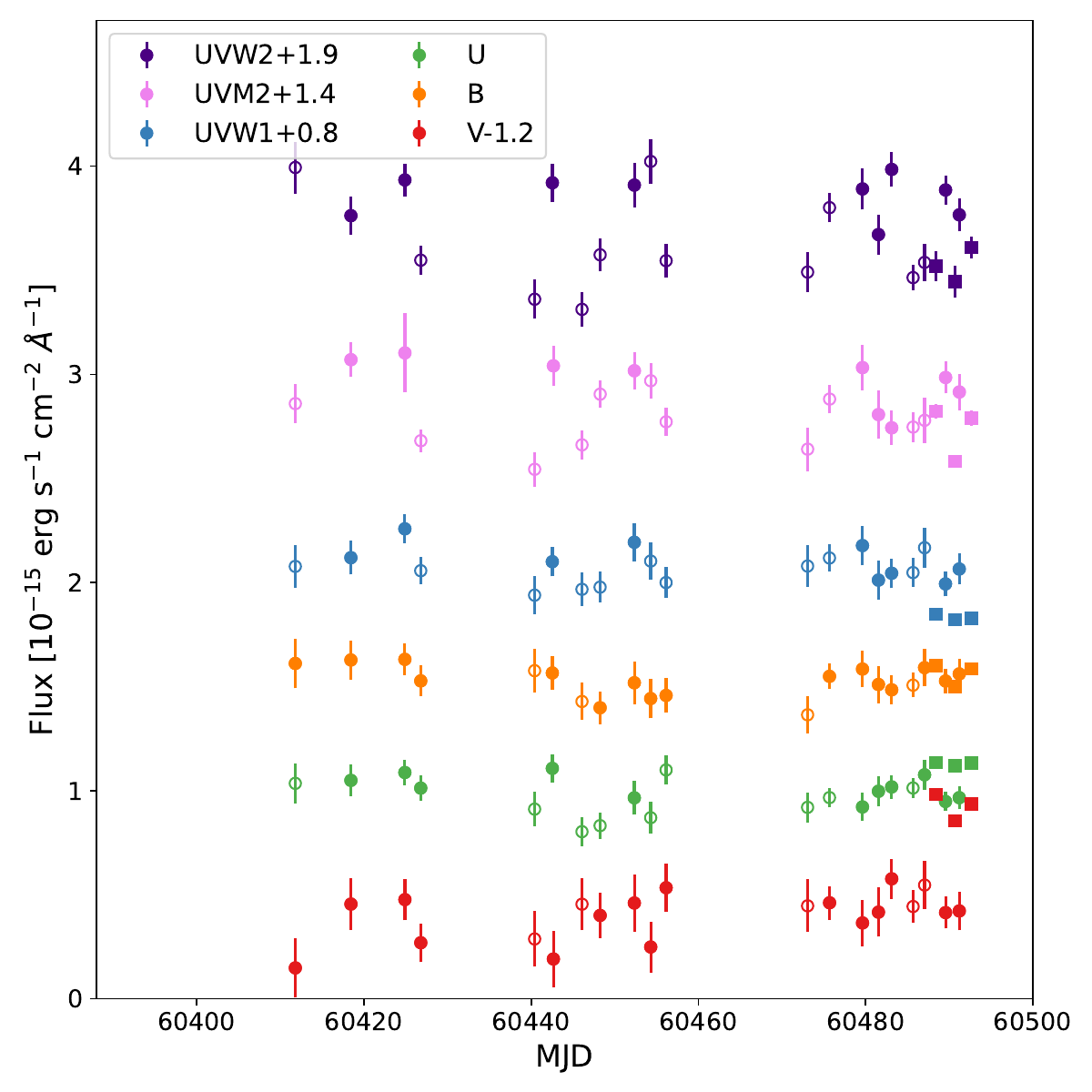}\\
 \end{tabular}
 
 \caption{\textit{Swift}/UVOT light curves. \textit{Left:} Full \textit{Swift}/UVOT light curves, constructed using data collected from July 31st, 2021 until July 10, 2024, including the UVW2 (purple circles), UVM2 (piink circles), UVW1 (blue circles), U (yellow circles), B (green circles), and V (red circles) bands. The period with gyroscope issues is shown in lighter colors, and observations in the low sensitivity areas of the detector are marked as empty symbols. The dashed lines show the results of the power-law decay fit for the UV bands. We obtain power-law indices of -0.23, -0.18, and -0.11, for the UVW2, UVM2, and UVW1 band, respectively. \textit{Right:} Zoom of the \textit{Swift}/UVOT light curves for the period between April 10 and July 10, 2024. As a reference, we show the \textit{XMM-Newton}/OM fluxes in squares. Note that the transmission curves of \textit{Swift}/UVOT and XMM/OM are not exactly the same.
 }
\label{fig:uvot_lc}
\end{figure*}

The left panel Extended Data Fig.~\ref{fig:uvot_lc} shows the full \textit{Swift}/UVOT flux light curves in the six filters obtained between July 31st, 2021, and July 10th, 2024. The season affected by the spacecraft jitter is shown in a lighter color, and the observations located in the low sensitivity areas are marked as empty symbols. In Fig.~\ref{fig:lc} we plotted only data not affected by these issues.
We modelled the UV filters with a power-law decay model in order to check if the flux decay follows the expectations for classical TDEs. We obtained power-law indices ($F\propto t^p$) of -0.23 for the UVW2 band, -0.18 for the UVM2 band and of -0.11 for the UVW1 band. The optical bands are dominated by the host flux, and thus, we did not model them. The indices obtained in the UV bands are similar to those reported in \cite{Sanchez24} for the difference flux in the ZTF $g$ and $r$ bands (-0.17 and -0.14, respectively), and are far from the theoretical predictions for classical TDEs (-5/3; \citeExtra{Rees88,Phinney89}), and from the values typically observed in optical TDEs ($-3.82 \lesssim p \lesssim -0.78$; \citeExtra{Hammerstein23}).

The right panel of Extended Data Fig.~\ref{fig:uvot_lc} shows the \textit{Swift}/UVOT light curves in the six filters between April 10 and July 10, 2024, i.e., dates not affected by spacecraft jitter.
The low number of data points not affected by the low sensitivity areas prevents from doing a variability analysis on timescales of days/weeks, which results in measurements of the order of 1$\sigma$ of confidence level.
For each of the observed bands, we estimated the normalized excess variance, $\sigma^{2}_{NXS}$ and its error $\Delta \sigma^{2}_{NXS}$, which represents the variability amplitude of the full light curves (corrected  for jitter and low throughput), and the confidence level of the variations, computed as the ratio between $\sigma^{2}_{NXS}$ and $\Delta \sigma^{2}_{NXS}$, following the prescriptions in \citeExtra{vaughan2003}.
Variations were detected in the UVW2 band ($\sigma^{2}_{NXS}$=0.022$\pm$0.004) and the UVM2 band ($\sigma^{2}_{NXS}$=0.023$\pm$0.005), where variations are detected at a 6$\sigma$ and 4$\sigma$ confidence level in timescales of years.
The other bands are more affected by the host galaxy flux, and thus, it is not surprising that none of them show variations with high confidence levels.

\subsubsection{\textit{XMM-Newton}/OM}

UV/optical luminosities were obtained from the optical monitor (OM) onboard {\textit{XMM-Newton} simultaneously with the X-ray data. 

The photometric filters are the following: UVW2 is centered at 1894 \AA\
(1805−2454) \AA, UVM2 at 2205 \AA\ (1970−2675) \AA, UVW1 at 2675 \AA\ (2410−3565) \AA, U at 3275 \AA\ (3030−3890) \AA, B at 4365	\AA\ (3719-4964) \AA, and V at 5438 \AA\ (4932-5963) \AA.
We retrieved the fluxes and AB magnitudes from the FITS source lists (OBSMLI)\footnote{https://xmmweb.esac.esa.int/docs/documents/XMM-SOC-GEN-ICD-0024.pdf}, which was produced by running the XMMSAS OM pipeline ({\sc omichain}). Extended Data Figure \ref{fig:uvot_lc} shows these fluxes with square markers. The differences in the XMM/OM and \textit{Swift}/UVOT fluxes are due to differences in the transmission filters of each instrument. 

We estimated the $\sigma^{2}_{NXS}$ for these bands and found that the UVM2 band has $\sigma^{2}_{NXS}$=0.009$\pm$ 	0.003, resulting in a significance of 3$\sigma$, thus rapid UV variations might be present in Ansky. 
The \textit{XMM-Newton} observations were obtained during the quiescence and QPE phases, thus it would remain to be determined if they are directly related to the QPEs.
If confirmed, these variations are unusual for a $\sim 10^6 M_\odot$ AGN, which should have a viscous time of $\mathcal{O}(10-100$) days; it is therefore possible that there is an additional driver of the rapid UV variability or that the central BH mass is much lower than what is predicted from the $M-\sigma$ scaling relations. UV variations will be studied in follow-up work.

\subsection{Radio data}

\subsubsection{ATCA}

In order to search for any radio emission associated with the transient, we observed the coordinates of Ansky on two occasions with the Australia Telescope Compact Array (ATCA; program: CX572). We observed the source on June 1, 2024, when the array was in the compact H168 configuration, and on July 27, 2024, when the array was in the compact H214 configuration. Both observations were taken with the CABB in the full 2048 spectral channel mode with the dual 5.5 and 9\,GHz dual receiver and 2\,GHz of bandwidth for each receiver. All data were reduced using standard procedures in the Common Astronomy Software Application \citeExtra[CASA][]{CASA2022}. PKS\,1934--638 was used for flux and bandpass calibration, and PKS\,1236+077 was used for secondary calibration. Two rounds of phase-only self-calibration were applied to the target field to aid with sidelobes in the target field due to a bright, nearby, AGN. Images of the target field were created using the CASA task \texttt{tclean}. The first observation resulted in a beam size of approximately 75x58$\arcsec$ at 5.5\,GHz and the second observation resulted in a beam size of approximately 56x28$\arcsec$ at 5.5 GHz. 

There is a compact radio source detected at a position of 13:35:18.93$\pm$5", 7:26:58.775$\pm$10", i.e. 1.18' away from Ansky and has a flux density of 342$\pm$60$\rm{\mu}$Jy at 5.5 GHz. This position is not consistent within error of the target and we deduce this radio source is likely unrelated to the target. At the location of the target there is no radio source detected in either of the observations at 5.5 or 9\,GHz, down to a 3$\sigma$ upper limit of $<$84$\rm{\mu}$Jy at 5.5\,GHz and $<$93$\rm{\mu}$Jy at 9\,GHz, corresponding to a 5.5\,GHz radio luminosity of $\nu L_{5.5\rm{GHz}}<6.5\times10^{39}$\,erg/s. 

\subsubsection{Radio surveys}
In addition to our dedicated ATCA observations, we investigated the presence of past or present radio activity through images from recent radio surveys. In particular, we considered the TIFR GMRT Sky Survey (TGSS, 150 MHz, \citeExtra{2017A&A...598A..78I}), the Rapid ASKAP
Continuum Survey (RACS-low, 0.8 GHz, and RACS-mid, 1.4 GHz, \citeExtra{2020PASA...37...48M}), and the VLA
Sky Survey (VLASS, 3 GHz, \citeExtra{2020PASP..132c5001L}). Frequencies below $\sim$1 GHz, tracing synchrotron emission from expanded, relaxed plasma, can provide information on the possible activation of a jet in previous epochs. For example, the extended lobes found in radio galaxies are more easily visible at these frequencies, given the enhanced flux due to their power-law spectrum (typically $S\propto\nu^{-1}$). In the MHz domain, the source results undetected in both TGSS and RACS-low, at a 3-$\sigma$level of 10 mJy and 1 mJy, respectively. At frequencies $>1$ GHz, where ongoing radio activation should be visible, the source was still undetected in RACS-mid in October 2020, with a 3-$\sigma$ upper limit of 0.54 mJy. As reported in \cite{Sanchez24}, the source was undetected in the first and second VLASS epochs, from April 2019 and November 2021, encompassing its activation period (December 2019). Subsequently, a third epoch was taken in July 2024, further confirming the non-detection (3-$\sigma$ upper limit 0.45 mJy). The TGSS and RACS-mid non-detection translates into a luminosity upper limit of $1.3\times10^{22}$ W/Hz at 150 MHz, while $7.1\times10^{20}$ W/Hz at 1.4 GHz. While the former would still be consistent with the low-luminosity tail of the local AGN population - as unveiled by the LoTSS survey at the same frequency \citeExtra{2019A&A...622A..17S} - the latter places the source more than 2 orders of magnitude below the conventional radio loud threshold ($10^{23}$ W/Hz \cite{1992ARA&A..30..575C}).

\section{Analysis}

\subsection{X-ray flare profiles}\label{sec:flare_profiles} To assess the peak-to-peak recurrence times, and determine the total energy output of each burst, we used an exponential rise/decay model which is common in the literature for QPEs \cite{Arcodia22,Chakraborty24}. The model has the form:
\[ \mathrm{QPE\;flux} = \begin{cases} 
      A\lambda e^{{\tau_1/(t_{peak}-t_{as}-t)}} & \mathrm{if}\; t<t_{peak} \\
      Ae^{{-(t-t_{peak})/\tau_2}} & \mathrm{if}\; t\geq t_{peak}
   \end{cases}
\]
where $A$ is the flare amplitude; $t_{peak}$ is the time of peak flux; $\tau_1$, $\tau_2$ are the e-folding times of rise and decay, respectively; $\lambda=e^{\sqrt{\tau_1/\tau_2}}$ is a normalization to join rise and decay; and $t_{as}=\sqrt{\tau_1 \tau_2}$ sets the asymptote time such that flux$=0$ for $t<t_{peak}-t_{as}$. The observed flares are generally well-described by this exponential profile (Fig.~\ref{fig:lc}), and as seen in other long-duration QPEs (e.g. eRO-QPE1, AT2019qiz), the rise is significantly more rapid than the decay.

As usually done in the literature, we define the flare duration as the interval encompassing $t_{\mathrm{peak}} - 3\tau_1 < t < t_{\mathrm{peak}} + 3\tau_2$, i.e. $\pm 3$ e-folds from the flare peak. 
This average is computed including the longer gaps between cascades, hence also the large standard deviation.
With these flare profiles, we find a mean peak-to-peak recurrence time of $t_{\mathrm{recur}} = 436.8\pm 80.3$~ks and mean duration of $t_{\mathrm{dur}} = 129\pm 43.2$~ks. The above definition of burst duration is used for our comparison to other QPEs in Fig.~\ref{fig:timescales}. We also computed the total energy outputs of the flares by integrating over the above definition of burst duration; we find an average integrated energy output of $(9.7\pm 3.7)\times 10^{47}$ erg s$^{-1}$.

The population of longer-period repeating transients from galactic nuclei is rapidly growing, at both optical \cite{Payne21} and X-ray \cite{Liu23,Evans23,Guolo24} wavelengths. There is large diversity and complexity in their multiwavelength properties, but generally, none of them show the same spectral evolution as in Ansky and other QPEs, suggesting a different emission mechanism. Their emission is likely powered by repeated accretion of stripped material from the outer layers of a captured star on an eccentric (but not deeply plunging) orbit, so that the captured EMRI survives several pericenter passages. In this sense, the longer-period transients probe a different region of the dynamical phase space in galactic nuclei; one particularly exciting science case connecting QPEs to the longer-period repeating nuclear transients is to constructing a coherent understanding of dynamical evolution in galactic nuclei, connecting the populations of long-period, high-eccentricity orbiters to the short-period and lower-eccentricity EMRIs at 100s of $R_g$.


\subsection{Probability of detecting QPEs before 2024}

X-ray emission was detected only after February 2024, while previous observations with \textit{Swift} and \textit{eROSITA} did not detect emission during the observations taken since December 2019. Since these observations were taken at random cadence, there exists a possibility that there was QPE emission before 2024, but that we did not observe it with the available observations.

The QPEs detected are all above the flux limit of the single observations of eROSITA ($\sim$1$\times10^{-13}$ erg s$^{-1}$ cm$^{-2}$), with a minimum flux measured in the NICER detections of 7.65$\times10^{-13}$ erg s$^{-1}$ cm$^{-2}$ and a maximum flux of 2.16$\times10^{-11}$ erg s$^{-1}$ cm$^{-2}$ in the $0.3-2$~keV band. Thus, we assume the only factor affecting the detection of QPEs like the ones observed in 2024 is related to the exact time of the eROSITA observations. To make a rough estimate of the probability of missing potential QPEs, we assumed a flare duration of 1.5 days and a peak-to-peak duration of 4.5 days. Then, given an observation period of 1.167 (or 0.833 or 0.999) days, we consider all possible starting times of both the event and the observation within their respective feasible intervals. By plotting these on a two-dimensional plane, we define a total area representing all possible combinations of start times. We then calculate the areas where the event and the observation do not overlap (where one ends before the other begins). This non-overlapping area represents the scenarios where the event is not detected during the observation. The probability of missing the event is the ratio of the non-overlapping area to the total area, resulting in approximately a 27\% (or 40\% or 33\%) chance of missing the event. Thus, the probability of missing a QPE in one single eRASS ranges between 27\% and 40\%, but we note this is a very broad estimate. Then, the combined probability of missing a flare in all five eRASS is of $\sim$0.26\%.
Considering, at the same time, that none of the \textit{Swift} observations prior to February 2024 detected a QPE, the probability of having missed prior activity, such as the one we see today, is quite low.

\section{Models \& Interpretation}

\subsection{Constraints on the system from the QPE timing}\label{sec:timing_models}

QPEs are usually modelled as stars or stellar-mass black holes orbiting around the SMBH (the so-called EMRIs) and crossing a pre-existing disk, producing two flares per orbit \citep{Xian21,Lu23,Franchini23,Linial23,Tagawa23}. 
The emission is produced by the shocked disk material during the EMRI--disk collisions, which gets hot enough to emit X-rays \citeExtra{Nayakshin2004}. 
This model reproduces the observed emission properties of QPEs presented in the literature \citep{Franchini23}, which Ansky broadly resembles.
The main differences are the longer duration and recurrence times.  The latter can be explained by an EMRI in a wider orbit as explained below.
The longer duration is a natural outcome if the disk density increases with radius, which is the case for a radiation-pressure dominated disk, as expected in this case \citep{Linial23}.
In this section we aim to reconcile the longer time-scales of the observed system with the EMRI model for QPEs.

If we simply consider the most common recurrence time observed for this source, the orbital period would be roughly 9 days, i.e., twice the observed period of 4.5 days. Given a SMBH mass of 10$^6\;M_\odot$, that implies an orbit semi-major axis for the EMRI of 8.5 au, roughly 850 times $R_{\rm g}$, the gravitational radius for the central black hole.  This is more than twice larger than all QPEs interpreted with this model \citep{Franchini23}. Moreover, that size sets a lower limit for the disk radius which corresponds to 18 times $R_{\rm t}$, the tidal disruption radius for a solar-type star. We can interpret this as further evidence for this source not being produced by a typical TDE, which expected circularization radius is $\sim 2 R_{\rm t}$, but perhaps by the SMBH accreting a gas cloud, which would be disrupted at a larger radius. 

A more compact disk could be responsible for the flares, provided that they are produced once per orbit.  In this case, the orbital period and semi-major axis for the EMRI would be 4.5 days and 5.3 au, respectively.
The smallest possible disk radius in this scenario would be achieved if the flares are produced at the EMRI's pericentre, i.e. if its orbit's semi-major axis is coplanar with the disk. 
Under that assumption, in order to have a disk with a radius smaller than $6 R_{\rm t}$, the eccentricity of the EMRI needs to be larger than 0.48. Such an EMRI would have a gravitational wave decay time of $\sim 2.5\times10^7\,$yr \citeExtra{Peters1964}, so it is not particularly unexpected to exist there.  However, relaxing any of the previous assumptions would result in a shorter decay time-scale.

This source has a more complex timing behaviour than what we just considered, with a longer, 6-day interruption every five flares, and flux variations on that same time-scale.  
We could envisage a situation in which the whole 25-day modulation is produced in a single orbit of a star or black hole crossing a complex (and fine-tuned) gaseous structure, formed by several evenly-spaced streams. That would require an even larger, $>16\,$au, size for the gas distribution.
Instead, in this section we consider extensions to the EMRI scenario.
Interruptions after every two flares are easily produced, provided the EMRI orbit is eccentric and crosses the disk twice per orbit, close to its pericentre.
Flux modulations are also reproduced, as precession of both the disk and the EMRI orbit change the location and relative velocity of the EMRI when crossing the disk \citep{Franchini23}.
The observed periodic interruptions after every five flares, however, seem harder to reproduce in this scenario, but it may involve an eccentric and/or warped disk, so precession results in some skipped collisions.
Schwarzschild precession of the EMRI orbit is likely too slow though.  In order to complete a precession cycle every six orbits, for a central mass of 10$^6\;M_\odot$ and a semi-major axis of 5.3 au, the required eccentricity is 0.983.  
For an EMRI mass of 10 M$_\odot$, the decay time-scale of such an orbit would be only 160 years \citeExtra{Peters1964}, which makes it unlikely to be present just at the time the AGN turned on.  Lense-Thirring precession would require an even higher eccentricity to be relevant.  

We can also consider the Schwarzschild precession of the disk, which has the advantage of not being tied to the observed flare recurrence.  
This precession can have observational consequences if the disk is eccentric. Such disks are expected to form after TDEs and likely also after the disruption of a gas cloud \citeExtra{Eracleous1996, Bonnell2008, Goicovic2016, Price2024}, and may take relatively long to circularise \citeExtra{Hayasaki2016}. More specifically, we consider a disk with an eccentric inner cavity and, for simplicity, an EMRI whose orbit does not precess and crosses the disk plane at its pericentre, every 4.5 days. Moreover, we require the disk cavity to have a precession period of 25 days, in order to match the observed time-scales of Ansky.  This results in a constrain for the orbital elements of the inner disk fluid elements, namely $a_{\rm d} = 1.33 {\rm au} (1-e^2_{\rm d})^{-2/5}$, where $a_{\rm d}$ and $e_{\rm d}$ stand for the inner disk semi-major axis and eccentricity, respectively. If the EMRI pericentre radius is in between the peri- and apo-centre of the disk inner cavity, then sometimes the pericentre passage will result in a flare, but sometimes it will not, as the EMRI will go through the precessing cavity.  These two conditions are plotted in Extended Data Figure~\ref{fig:ecc_cavity}, where we can see that there is a wide range of parameters that potentially can fit Ansky.  In particular, relatively low values for the disk and EMRI eccentricities, 0.4 and 0.7 respectively, appear as realistic options.  We notice that such an EMRI would have a relatively long decay time-scale of $3.7\times10^6\,$yr \citeExtra{Peters1964}.

We acknowledge that we have not presented an actual model to explain the peculiar timing behaviour of this QPE.  Doing so would require including the simultaneous relativistic precession of both the EMRI orbit and the disk, plus the hydrodynamical evolution of the latter, all of which is out of the scope of this study.  However, we have shown that the relevant time-scales in principle allow an explanation of this system as an EMRI interacting with an accretion  disk featuring a precessing, eccentric cavity.

\begin{figure}
\centering
\includegraphics[width=0.5\textwidth]{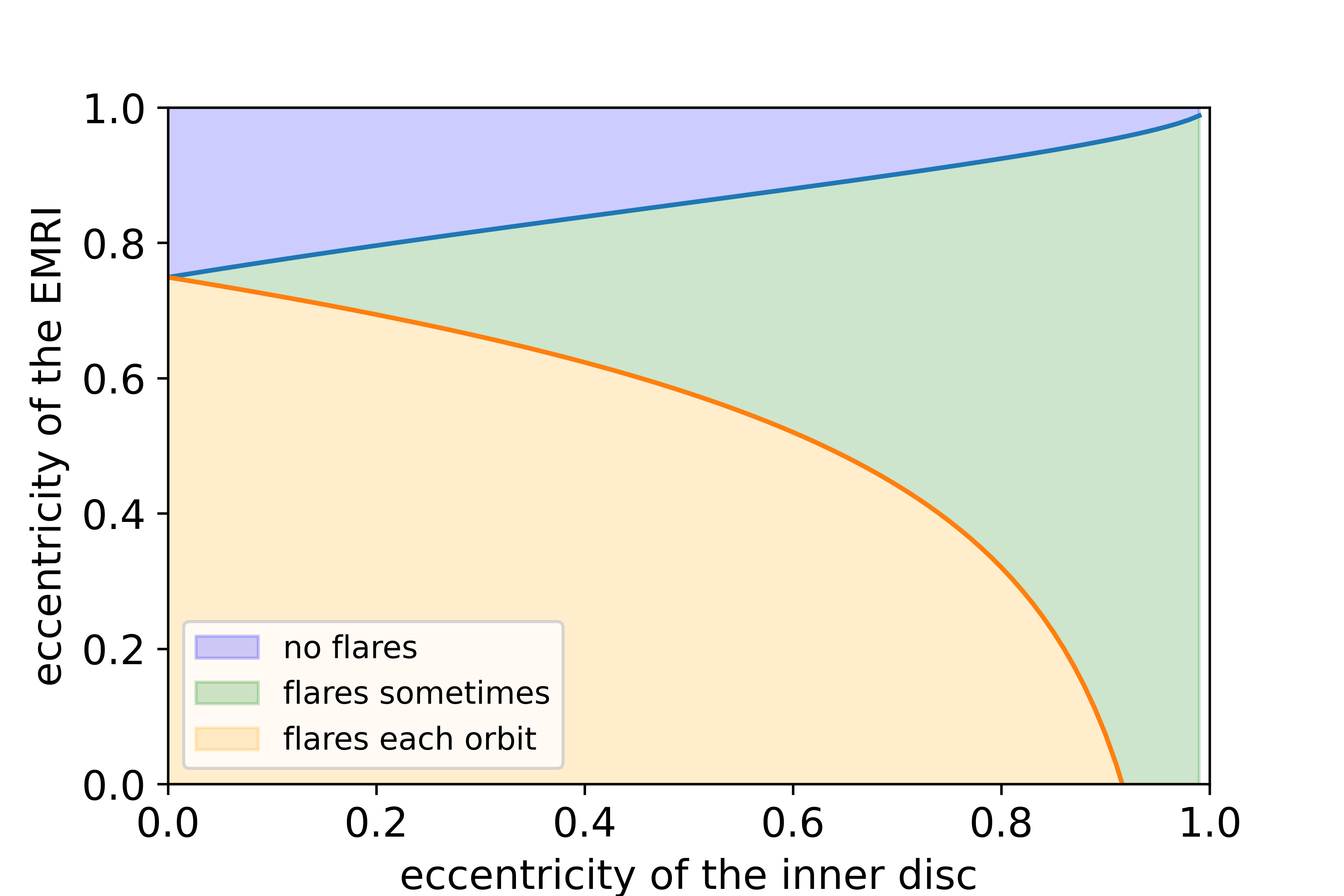}
\caption{Parameter space of disk and EMRI eccentricities for the case of an EMRI crossing the disk plane at pericentre. The orbital period of the EMRI is fixed by the observed QPE periodicity, while the properties of the inner disk are set so it completes a precession cycle every 25 days. 
The upper area represents cases in which the EMRI eccentricity is too high, resulting in a pericentre that always go through the cavity.
In the lower area the EMRI has a wider pericentre, and always crosses the disk.
In the middle area the EMRI sometimes crosses the disk and sometimes goes through the cavity, potentially explaining the behaviour observed in Ansky.
}
\label{fig:ecc_cavity}
\end{figure}

\subsection{Constraints on the system from the optical and UV fluxes}

We now consider whether the disk required to produce the QPE flares is consistent with the baseline optical and UV emission of the source after it `turned on'.  Assuming a standard accretion disk model \citeExtra{shakura1973}, we estimate the monochromatic flux density produced up to different outer truncation radii for different wavelengths. We then compare these values to the measured fluxes, using a luminosity distance of 105 Mpc. We note that from Fig. 4 in \cite{Sanchez24}, the flux in the UV-bands W2 ( $\lambda_{eff}$ = 2052\AA), M2 ($\lambda_{eff}$ = 2245\AA) and W1 ($\lambda_{eff}$ = 2682\AA) in this object is four or more times brighter after the ignition event than before, therefore the \textit{Swift} measurements in these bands are only slightly contaminated by the host galaxy and most of the flux could be attributed to a newly formed accretion disk. From the same figure, it is clear that longer wavelength bands are heavily contaminated by the host galaxy. Therefore, to obtain a clean $g$-band flux that could be attributed to an accretion disk, we subtracted the mean pre-ignition flux, measured from the ZTF lightcurve, from a post-ignition flux measured at a date close to the \textit{Swift} observation used for the other fluxes.  These observed values are shown as horizontal dashed lines in Extended Data Fig.~\ref{fig:disc_model}. 

Extended Data Fig.~\ref{fig:disc_model} (top panel) shows that a standard disk with an accretion rate corresponding to an Eddington ratio of unity and a black hole mass of $10^6\;M_\odot$ matches the observed flux of the shortest wavelength UV band when integrated up to a radius of 341 R$_g$.  For these accretion disk parameters, the other UV bands fulfil their observed fluxes at similar truncation radii of 373 and 469 R$_g$.  Note however that the redder UV bands should have some host-galaxy contamination, making their flux slightly higher and therefore requiring integration out to a larger radius.  We can therefore conclude that the UV fluxes are consistent with a disk of outer radius $\approx$ 350 R$_g$, which would match the estimates of the previous section in the case of an eccentric EMRI orbit.  
The disk size is also consistent with the interpretation of previous QPEs that require EMRI orbits of semi-major axes up to 355 R$_g$, and therefore disks of that scale \citep{Franchini23}.  While such a disk would be a few times more extended than the circularisation radius expected for the TDE debris, viscosity will spread the material and may allow it to reach those sizes after a few years of evolution \citep{Franchini23}.
On the other hand, the optical $g$ band flux requires integration out to a much larger radius of 756 R$_g$.  
Another component is then required to contribute to the optical emission, such as the optically-thick envelope found in recent TDE simulations \citeExtra{Price2024}. 
All in all, the QPE properties and observed UV and optical fluxes seem consistent with the aftermath of a standard TDE, despite its extreme inferred properties \cite{Sanchez24}, but only if the additional optically thick envelope is included in the model.

Alternatively, a match for both the UV and optical fluxes with a single component can be obtained by assuming a disk with lower accretion rate but larger outer radius, consistent with the idea of a disrupted cloud rather than a standard TDE, as discussed in the previous section.   Extended Data Fig.~\ref{fig:disc_model} (bottom panel) shows as an example the expected emission from a disk with an Eddington ratio of 0.28. In this case, the fluxes of all 4 bands can be achieved, within the uncertainties, with a single accretion disc truncated at $\sim 1200$ R$_g$. This radius is three times larger than those predicted for standard TDEs, thus it would be more consistent with the hypothesis of a turn-on AGN or of an exotic nuclear transient.

\begin{figure}
\centering
\includegraphics[width=0.5\textwidth]{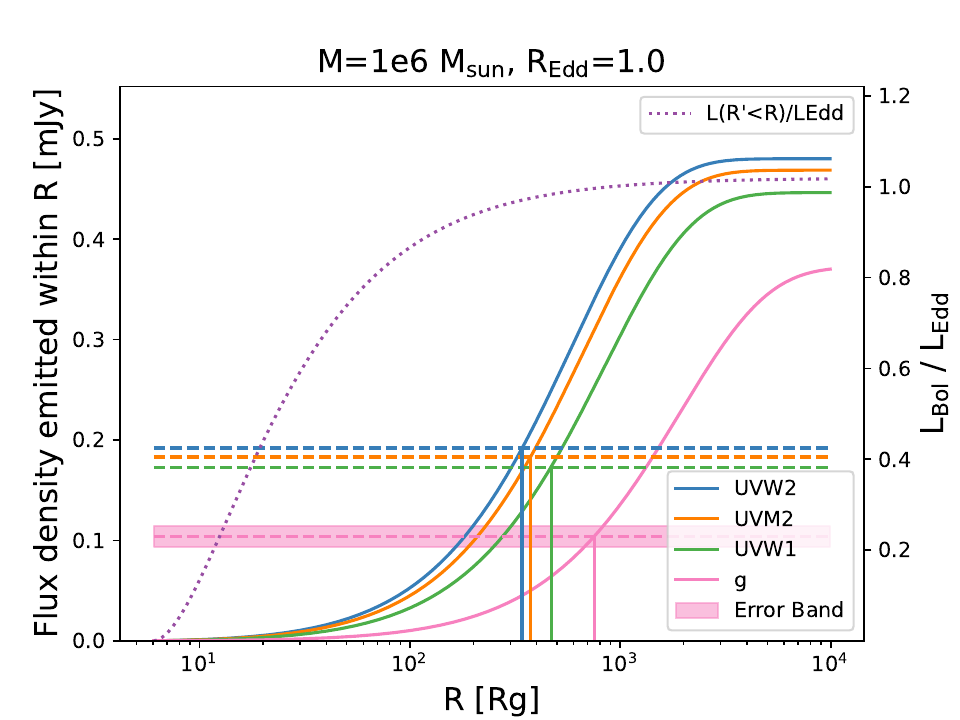}
\includegraphics[width=0.5\textwidth]{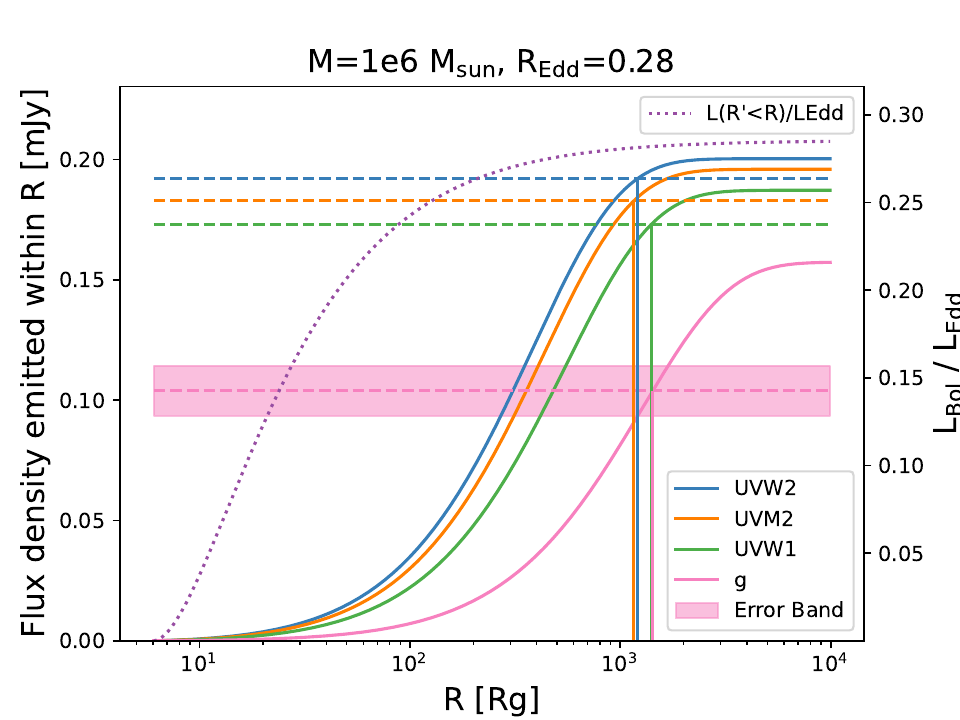}
\caption{The solid lines show the cumulative flux density emitted by a standard accretion disk as a function of outer truncation radius, for four different wavelengths. The dashed lines of the corresponding colors show the measured flux densities for each band and the vertical lines mark the truncation radii needed to satisfy each observed flux. The dotted line shows the cumulative bolometric luminosity integrated from the same disk model as a function of outer truncation radius, divided by the Eddington luminosity. The error band marks the effect of a 10\% uncertainty in the g-band band flux, for this non-simultaneous observation. In the top panel the accretion rate is set to match the UVW2 flux at a truncation radius of 330 $R_g$, expected for the size of a TDE disc. This truncation radius under-predicts the optical flux by a factor of 2. In the bottom panel the truncation radius and accretion rates are fit to match the measured fluxes in the optical and UV bands simultaneously. Notice the different dynamical range of the y-axis on both plots.
}
\label{fig:disc_model}
\end{figure}

\bibliographyExtra{main}

\end{document}